\documentclass[10pt]{iopart}
\pdfoutput=1
\usepackage[top=1.2 in, bottom=0.75 in, left=0.9 in, right=0.9 in]{geometry}

\usepackage{footmisc}
\usepackage{iopams}
\expandafter\let\csname equation*\endcsname\relax
\expandafter\let\csname endequation*\endcsname\relax
\usepackage{amsmath,amssymb,amsfonts}
\usepackage{graphicx,float,color}
\usepackage{bbold,bbm}
\usepackage{tikz,drawInt}
\usepackage[colorlinks,bookmarks=false,citecolor=blue,linkcolor=blue,hyperfootnotes=true,urlcolor=blue]{hyperref}

\newcommand{\be}{\begin{equation}}
\newcommand{\ee}{\end{equation}}
\newcommand{\bea}{\begin{eqnarray}}
\newcommand{\eea}{\end{eqnarray}}

\newcommand{\ra}{\rangle}
\newcommand{\dg}{^\dagger}
\newcommand{\p}{\partial}
\newcommand{\rd}{{\rm d}}

\newcommand{\bw}{\begin{widetext}}
\newcommand{\ew}{\end{widetext}}

\newcommand{\sech}{{\rm sech}}

\def\nn{\nonumber\\}

\def\fr#1{(\ref{#1})}

\usepackage{etoolbox}
\makeatletter
\def\@mkboth#1#2{}
\newlength\appendixwidth
\preto\appendix{\addtocontents{toc}{\protect\patchl@section}}
\newcommand{\patchl@section}{%
  \settowidth{\appendixwidth}{\textbf{Appendix }}%
  \addtolength{\appendixwidth}{1.5em}%
  \patchcmd{\l@section}{1.5em}{\appendixwidth}{}{\ddt}%
}
\makeatother

\makeatletter
\def\footnoterule{\kern-3\p@
  \hrule \@width 2in \kern 2.6\p@} 
\makeatother

\begin{document}

\title{Excitations in the Yang-Gaudin Bose gas}

\author{Neil J Robinson}
\address{Condensed Matter Physics and Materials Science Division, \\Brookhaven National Laboratory, Upton, NY 11973-5000, USA\\
E-mail: \href{mailto:nrobinson@bnl.gov}{nrobinson@bnl.gov}}

\author{Robert M Konik}
\address{Condensed Matter Physics and Materials Science Division, \\Brookhaven National Laboratory, Upton, NY 11973-5000, USA\\
E-mail: \href{mailto:rmk@bnl.gov}{rmk@bnl.gov}}

\date{\today}

\begin{abstract}
We study the excitation spectrum of two-component delta-function interacting bosons confined to a single spatial dimension, the Yang-Gaudin Bose gas. We show that there are pronounced finite-size effects in the dispersion relations of excitations, perhaps best illustrated by the spinon single particle dispersion which exhibits a gap at $2k_F$ and a finite-momentum roton-like minimum. Such features occur at energies far above the finite volume excitation gap, vanish slowly as $1/L$ for fixed spinon number, and can persist to the thermodynamic limit at fixed spinon density. Features such as the $2k_F$ gap also persist to multi-particle excitation continua. Our results show that \textit{excitations in the finite system can behave in a qualitatively different manner to analogous excitations in the thermodynamic limit}.

The Yang-Gaudin Bose gas is also host to multi-spinon bound states, known as $\Lambda$-strings. We study these excitations both in the thermodynamic limit under the string hypothesis and in finite size systems where string deviations are taken into account. In the zero-temperature limit we present a simple relation between the length $n$ $\Lambda$-string dressed energies $\epsilon_n(\lambda)$ and the dressed energy $\epsilon(k)$. We solve the Yang-Yang-Takahashi equations numerically and compare to the analytical solution obtained under the strong couple expansion, revealing that the length $n$ $\Lambda$-string dressed energy is Lorentzian over a wide range of real string centers $\lambda$ in the vicinity of $\lambda = 0$. We then examine the finite size effects present in the dispersion of the two-spinon bound states by numerically solving the Bethe ansatz equations with string deviations.  

$ $

\end{abstract}

\maketitle

\hrule height 1.5pt

\tableofcontents
\vspace{0.8cm}
\hrule height 1.5pt

\section{Introduction}

\subsection{Background}

Quantum integrable models provide firm ground from which one can gain understanding of the physics of strongly correlated systems. They include paradigmatic examples of magnetism (the spin-$1/2$ Heisenberg model~\cite{BetheZPhys31,TakahashiBook}), interacting electrons on the lattice (the Hubbard model~\cite{LiebPRL68,HubbardBook}), and interacting Bose and Fermi gases (the Lieb-Liniger~\cite{LiebPR63a}--\cite{KorepinBook} and Yang-Gaudin~\cite{YangPRL67}--\cite{GaudinBook} models). As a result of their exact solutions, integrable models can be used to study the non-perturbative effects of interactions on quasi-particle excitations, hopefully paving the way for insights in more generic cases. 

As a starting point for attacking problems with strong correlations, it is clearly desirable to understand the nature and characteristics of quasi-particle excitations in quantum integrable models. By now there is a well-trodden path for such studies: the logarithmic Bethe ansatz equations suggest a natural definition of excitations in terms of the Bethe roots (and their defining integer quantum numbers)~\cite{BetheZPhys31}. One can then extract the dispersion relation for single-particle excitations by direct numerical computation in finite-size systems, perturbative calculations in the weak or strong coupling limits, or by working directly in the thermodynamic limit and performing manipulations of systems of coupled integral equations. The quasi-particle excitations can be combined to describe the continua of multi-particle excitations. Furthermore, the Bethe ansatz equations also provide a description of emergent multi-particle excitations, such as bound states, which are characterized by complex Bethe roots~\cite{BetheZPhys31,TakahashiBook,HubbardBook,KorepinBook}. The presence of such bound states is model dependent (for example, there are no multi-particle bound states in the repulsive Lieb-Liniger model~\cite{LiebPR63a}--\cite{KorepinBook}, although they are present in the attractive limit~\cite{KorepinBook}).  There may be multiple types of bound states, such as the well known $k$-$\Lambda$ and $\Lambda$ strings in the one-dimensional Hubbard model~\cite{HubbardBook}.

Except for some special cases, numerical computations are usually a necessity -- especially in the case of multicomponent systems, where the system of Bethe ansatz equations is nested (e.g., there are additional sets of auxiliary Bethe roots), see for example Ref.~\cite{LiEPL03}. Often such calculations are performed on a finite-size systems, so it is useful to understand how the properties of solutions change with the system size (so-called \textit{finite-size effects}). The study of finite-size effects can be a useful theoretical tool in the study of critical phenomena, allowing one to extract the central charge~\cite{BlotePRL86,AffleckPRL86}, critical exponents~\cite{CardyJPhysA84}--\cite{MironovPRL91} and the operator content~\cite{LauchliArxiv13,SchulerArxiv16} of conformal field theories. They also have extensive applications in integrable quantum field theories~\cite{RavaniniArxiv01} and have recently received attention for the computation of threshold singularities in integrable lattice models~\cite{PereiraPRL08}--\cite{VenessPRB16}. Of course, understanding the properties of finite-size systems is also useful when comparing the result of theoretical calculations to experiments on intrinsically finite-size systems.

Interest in quantum integrable models has recently undergone a resurgence (see, for example, the recent reviews~\cite{EsslerJStatMech16}--\cite{DeLucaJStatMech16}), thanks to ground-breaking progress in the field of ultra-cold atomic gases~\cite{LewensteinAdvPhys07}. Experiments have achieved both unprecedented levels of isolation from the environment and control of the Hamiltonian, which have lead to extremely accurate realizations of oft-studied theoretical models~\cite{LewensteinAdvPhys07}. Thanks to this, integrable quantum systems are now routinely studied in the laboratory\footnote{To be more precise, the experiments on cold atomic gases are weakly non-integrable, in the sense that integrability is (weakly) broken by small inhomogeneities, the presence of a trapping potential, boundary conditions, and so forth. However, it has been understood theoretically that such weak breaking of integrability can lead to dynamics that remain approximately integrable for very long periods of time, see for example, Refs.~\cite{LangenJStatMech16},\cite{MoeckelPRL08}--\cite{BertiniArxiv16}.}, with both their equilibrium properties~\cite{LewensteinAdvPhys07,StamperKurnRMP13,GuanRMP13} and their non-equilibrium dynamics~\cite{PolkovnikovRMP11}--\cite{AlessioArxiv15} receiving a great deal of scrutiny. In the latter case, the ground-breaking experiments of Kinoshita, Wenger and Weiss~\cite{KinoshitaNature06} highlighted the dramatic consequences of integrability on non-equilibrium dynamics; integrable systems driven out of equilibrium do not thermalize, but instead equilibrate to a \textit{generalized Gibbs ensemble} whose form is fixed by the initial expectation values of local and quasi-local conservation laws~\cite{RigolNature08,IlievskiPRL15}. 

\subsection{This work}

In this work we study properties of the excitations in the Yang-Gaudin Bose gas, with a particular focus on finite-size effects and the multi-particle bound states. In the first part of this paper, we will focus our attention on the spinon $(s)$ dispersion, and the two-particle continua for holon-antiholon ($h\bar h$) and spinon-holon ($sh$) excitations. These will be studied using both the exact numerical solution of the Bethe ansatz equations and the strong coupling expansion. Whilst these excitations have previously received some attention~\cite{LiEPL03}, only small finite systems were considered and there was no study of finite-size effects. \emph{We will show that the finite-size effects in this system are large and quite surprising: excitations in the finite system can behave in a qualitatively different manner to analogous excitations in the thermodynamic limit}. This is particularly well illustrated by the spinon dispersion, which exhibits a pronounced finite-momentum roton-like minima in small systems which vanishes in the thermodynamic limit ($M/L \to 0$, with $M$ the number of spinons). We study how such features evolve with system size and present numerical evidence that the finite-momentum roton-like minima can persist to the thermodynamic limit of the two-component gas provided the spinon density does not vanish ($M, N, L \to \infty$ with $N/L$, $M/L$ fixed, $N$ being the number of particles). 

Our interest in understanding the simple few-particle excitations of the Yang-Gaudin Bose gas stems from recent works on the non-equilibrium dynamics of a distinguishable impurity in the Bose gas~\cite{RobinsonPRL16,RobinsonArxiv16}. The dynamics of an impurity exhibit some interesting features: an initially localized impurity can undergo arrested expansion or may move through the gas in a snaking motion~\cite{RobinsonPRL16}. Furthermore, for initial states containing a superposition of spinon excitations the spreading shows clear signs of a `double light cone', which is related to the presence of a roton-like minima in the spinon dispersion~\cite{RobinsonArxiv16}. As the impurity limit of the Yang-Gaudin Bose gas ($M=1$, $N/L$ finite) is perhaps the simplest limit to consider and already exhibits unusual non-equilibrium behavior, it is necessary to develop a good understanding of the excitations of the model. There has also been interesting recent work which relates the non-relativistic limit of various integrable relativistic quantum field theories to multi-component Lieb-Liniger and Yang-Gaudin models~\cite{BastianelloJStatMech16,BastianelloArxiv17}. 

In the remainder of the paper, we turn our attention to the multi-spinon bound states present in the Yang-Gaudin Bose gas. Starting in the thermodynamic limit, we numerically solve the Yang-Yang-Takahashi equations to compute the dispersion relation for bound states, so-called $\Lambda$-strings. We present a simple relation between the dressed energies of the $\Lambda$-strings and antiholon dressed energy $\epsilon(k)$, which we solve under a strong coupling expansion to reveal a particularly simple closed form. We then consider finite size effects for the bound state excitations in small systems, taking into account string deviations. 

We finish with a discussion of our results, including implications for comparison between theory and experiments in cold atomic gases.

\subsection{The Yang-Gaudin Bose gas}

The Yang-Gaudin Bose gas\footnote{This is often also called the two-component Lieb-Liniger model or the spinor (two-component) Bose gas. The Yang-Gaudin model can also refer to the same Hamiltonian with spin-$1/2$ fermionic fields. Here we will only discuss two-component bosons.} is described by the Hamiltonian density
\bea
{\cal H} =\frac{\hbar^2}{2m} \sum_{j=1,2} \p_x \Psi\dg_j(x) \p_x \Psi_j(x) + c \sum_{j,l=1,2}  \Psi\dg_j(x) \Psi\dg_l(x) \Psi_l(x) \Psi_j(x), 
\eea
where $j,l=1,2$ label the two different boson species, $m$ is the boson mass, and $c$ characterizes the interaction strength. The bosonic fields obey canonical commutation relations,
\be
\Big[\Psi_j(x), \Psi\dg_l(y)\Big] = \delta_{j,l} \delta(x-y).
\ee
Herein we set $\hbar = 2m = 1$. This model is integrable, and may be solved via the nested Bethe ansatz~\cite{YangPRL67}--\cite{GaudinBook}. The $N$ particle eigenstates, with $M$ particles of the second species, are described by two sets of quantum numbers: the momenta $\{k\} \equiv \{ k_1,\ldots, k_N\}$ and the spin rapidities $\{ \Lambda \} \equiv \{ \Lambda_1, \ldots, \Lambda_{M}\}$. The spin rapidities are often known as auxiliary Bethe roots, as they do not directly enter into expressions for the momentum or energy of the eigenstates, see Eqs.~\fr{Eq:P}~and~\fr{Eq:E}. The quantum numbers satisfy the \textit{Bethe ansatz equations}, which read in their logarithmic form~\cite{YangPRL67}--\cite{SutherlandPRL68}
\bea
2 \pi I_j &=& k_j L + \sum_{l=1}^N \phi_1 (k_j, k_l) - \sum_{\beta = 1}^{M} \phi_2 (k_j, \Lambda_\beta), \label{FullBA1}\\
2\pi  J_\alpha &=& \sum_{l=1}^N \phi_2 (\Lambda_\alpha ,  k_l) - \sum_{\beta = 1}^{M} \phi_1(\Lambda_\alpha , \Lambda_\beta). \label{FullBA2}
\eea
Here we have defined the scattering phase 
\be
\phi_n(u,v) = \rmi \log\left( \frac{\rmi c + (u-v)n}{\rmi c - (u-v)n} \right) \equiv 2 \arctan\left(\frac{(u-v)n}{c} \right), \label{Eq:phase}
\ee
and the sets of `integers' $I_j,\ J_\alpha$ which obey 
\bea
I_j,  J_\alpha \in  \left\{ \begin{array}{lcl} \mathbbm{Z}, & \quad & \mathrm{if~} (N+M) \in 2\mathbbm{Z}+1,\\
\mathbbm{Z} + \frac12, & \quad & \mathrm{if~} (N+M)\in 2\mathbbm{Z}. \end{array} \right. \label{Eq:integers}
\eea
The eigenstate $|\{k\};\{\Lambda\}\ra$ associated with the sets of integers $\{I\}$, $\{J\}$ has momentum $P$ and energy $E$ given by
\bea
P(\{I\},\{J\}) &=& \sum_j k_j = \frac{2\pi}{L}\left( \sum_j I_j - \sum_\beta J_\beta \right), \label{Eq:P}\\
E(\{I\},\{J\}) &=& \sum_j k_j^2.  \label{Eq:E}
\eea

The zero-temperature ground state of the Yang-Gaudin Bose gas is fully polarized: it is found in the sector with $M = 0$, which follows from general symmetry considerations~\cite{SutoJPhysA93,EisenbergPRL02}.\footnote{A system of multi-component interacting bosons with component-independent repulsive interaction will have a ferromagnetic ground state~\cite{SutoJPhysA93,EisenbergPRL02}.} With $N$ even (and herein we take $N$ to be even), the $N$-particle ground state is described by the set of momenta integers
\be
I_{\rm GS} = \left\{ -\frac{N-1}{2}, -\frac{N-1}{2}+1,\ldots, \frac{N-1}{2} \right\}, \label{Eq:IntegersGS}
\ee
which forms a `Fermi sea' about the origin. As the ground state is fully polarized, it coincides with the ground state of the Lieb-Liniger model for a single component Bose gas~\cite{LiebPR63a}--\cite{KorepinBook}. 

\subsection{Low-energy excitations of the Yang-Gaudin Bose gas}
\label{excitations}

The Bethe ansatz equations~\fr{FullBA1},\fr{FullBA2} and the integers~\fr{Eq:integers} provide a natural definition for excitations. Restricting our attention to states containing $N$ particles, and starting from the absolute ground state~\fr{Eq:IntegersGS}, we can construct the following types of excitations~\cite{LiEPL03}:
\begin{enumerate}
\item {\bf The spinon ($s$) excitation.} This corresponds to the presence of a single spin rapidity $\Lambda$ in the Bethe ansatz equations~\fr{FullBA1},\fr{FullBA2}, characterized by the integer $J_s$. According to the rules~\fr{Eq:integers}, the momenta integers shift and we have the following configuration
\bea
I_s = \left\{ -\frac{N}{2}, \ldots, \frac{N}{2}-1 \right\}, \qquad  | J_s| \leq \frac{N}{2}.  \label{Eq:Is}
\eea  
This corresponds to a `spin wave' excitation~\cite{FuchsPRL05,ZvonarevPRL07}, where the species index plays the role of spin (accordingly, this excitation is sometimes called an \textit{isospinon}). We show an example configuration for $N=8$ particles below:
\begin{center}
        \integers
        \border
        \intrange{-9}{8}
        \fillI{-4}{3}
        \intLabel{I}
        \render
\end{center}
\begin{center}
        \integers
        \intrange{-4}{4}
        \fillI{-2}{-2}
        \particleadd{-2}{J_s}
        \intLabel{J}
        \render
\end{center}

\item {\bf The holon-antiholon ($h\bar h$) excitation.}  Here we start from the ground state configuration of integers~\fr{Eq:IntegersGS} and remove the $j$th integer [leaving a single hole at $-(N-1)/2+j$]. We then add an integer $I_N$ outside the Fermi sea:
\bea
I_{h\bar h} = \left\{ -\frac{N-1}{2}, \ldots, -\frac{N-1}{2}+j-1, -\frac{N-1}{2}+j + 1,\ldots, \frac{N-1}{2}, I_N \right\}, \nn
{\rm with~~} I_N  > \frac{N-1}{2} {\rm ~~and~~} 1 \leq j \le N.
\eea
We illustrate this configuration (with $I_h = j - (N-1)/2$ the position of the hole in the sea of integers) for $N=8$ particles below:
\begin{center}
        \halfintegers
        \border
        \intrange{-7}{8}
        \fillI{-3}{4}
        \holeadd{1}{I_h}
        \particleadd{7}{I_N}
        \intLabel{I_\alpha^1}
        \render
\end{center}

\item {\bf The spinon-holon ($sh$) excitation}. This two-particle excitation is similar to the combination of the above two excitations. We consider the configuration of integers,   
\bea
I_{sh} = \left\{ -\frac{N}{2}, \ldots, -\frac{N}{2}+j-1, -\frac{N}{2}+j + 1,\ldots, \frac{N}{2} \right\}, \nn
J_{sh} \leq \frac{N}{2}, \qquad {\rm and}\qquad 1 \le j < N, \label{Eq:IntegersSH}
\eea
i.e., we consider a symmetric Fermi sea of integers containing a single hole at position $j$ accompanied by a single spin rapidity. For $N=8$ particles this configuration has the following diagrammatic depiction:
\begin{center}
        \halfintegers
        \border
        \intrange{-7}{8}
        \fillI{-3}{4}
        \holeadd{1}{I_h'}
        \particleadd{7}{I_N}
        \intLabel{I}
        \render
\end{center}
\begin{center}
        \integers
        \intrange{-4}{4}
        \fillI{-2}{-2}
        \particleadd{-2}{J_{sh}}
        \intLabel{J}
        \render
\end{center}
\noindent where $I_h' = -N/2+j$ is the position of the hole in the Fermi sea of integers.

\item {\bf Spinon bound states ($\Lambda$-strings).} When the system contains more than one flipped spin the associated spin rapidities can become complex. Such solutions are arranged in regular patterns in the complex plane, known as strings~\cite{BetheZPhys31,TakahashiPTP71}. An $n$-string $\Lambda^{(n)}_\alpha \equiv \{\Lambda^{n,1}_\alpha,\ldots,\Lambda^{n,n}_\alpha\}$ consists of $n$ spin rapidities which share the same real part $\omega_\alpha^n$ 
\be
\Lambda_\alpha^{n,a} = \omega_\alpha^n + i \frac{c}{2}(n+1-2a) + i\zeta_{\alpha}^{n,a}.
\ee
$\zeta_{\alpha}^{n,a} \in \mathbbm{C}$ are known as the `string deviations', which are non-zero in the finite-size system. It is usually assumed that the string deviations are exponentially small in the system size $L$, and in the thermodynamic limit can be set to $\zeta_{\alpha}^{n,a} = 0$; this is known as the `string hypothesis'. Rapidities in each string are then spaced evenly in the complex plane. 

\end{enumerate}

\section{The spinon single-particle dispersion}
\label{Sec:Spinon}

We consider the case with $N$ particles, of which there is a single $M = 1$ particle of the second kind. In this case, the Bethe ansatz equations~\fr{FullBA1} and~\fr{FullBA2} become particularly simple~\cite{PozsgayJPhysA12}. In their logarithmic form they read
\bea
2 \pi I_j &=& k_j L + \sum_{l=1}^N \phi_1 (k_j , k_l) - \phi_2 ( k_j , \Lambda), \label{BA1} \\
2 \pi J &=& \sum_{l=1}^N \phi_2(\Lambda , k_l), \label{BA2}
\eea
where the `integers' $I_j,\ J$ satisfy Eq.~\fr{Eq:integers} and $|J| \le N/2$, which follows from the bounding of $\phi_2(u)$. The momentum and energy of the eigenstates are as previously described in Eqs.~\fr{Eq:P} and~\fr{Eq:E}, respectively. 

Herein we focus on the case with $N\in 2\mathbbm{Z}$. We choose conventions where the ground state in the sector with $M=1$ is described by the integers. 
\be
I_0 \equiv \left\{ -\frac{N}{2}, -\frac{N}{2}+1, \ldots, \frac{N}{2}-1 \right\}, \qquad J_0 = -\frac{N}{2}, \label{Eq:GSconfig}
\ee
and we exclude $J = N/2$ from future discussions, to avoid double counting. Diagrammatically, for $N=8$ particles this is:
\begin{center}
        \integers
        \border
        \intrange{-9}{8}
        \fillI{-4}{3}
        \intLabel{I_0}
        \render
\end{center}
\begin{center}
        \integers
        \intrange{-4}{3}
        \fillI{-4}{-4}
        \particleadd{-4}{J_0}
        \intLabel{J}
        \render
\end{center}
Notice that when $J > -N/2$, it follows from Eq.~\fr{Eq:P} that the state has finite-momentum and the left/right Fermi points of the set of momenta $\{k_j\}$ do not coincide. 

\subsection{Strong coupling expansion}

To compute the spin wave dispersion under a strong coupling ($1/c$) expansion, we follow the standard prescription described in, e.g., Ref.~\cite{KorepinBook}. Our aim is to compute the energy associated with the presence of a spin wave (e.g., the presence of a spin rapidity $\Lambda$, or integer $J$) in the system. We compute the energy of the spin wave state $J > -N/2$ above the ground state configuration~\fr{Eq:GSconfig} with the set of integers $I_0$ fixed. 

\subsubsection{Integral equation for the shift function.} 

In order to compute the spin wave dispersion above the ground state, we fix the integers $I$ in the ground state configuration $I_0$ and vary the integer $J$ from its ground state value $J_0 = -N/2$ to $J > -N/2$ (the ground state value of the spin rapidity is $\Lambda_0 = -\infty$, which corresponds to acting on a Lieb-Liniger eigenstate with a global spin lowering operator~\cite{PozsgayJPhysA12}). We proceed by taking the difference of the first Bethe equation for the two cases; we denote the ground state momenta by $\{k_j^{(0)}\}$ and the excited state momenta by $\{k_j\}$ with the accompanying finite spin rapidity $\Lambda$. We have  
\bea
0 &=& \Big( k_j - k_j^{(0)} \Big)L - \phi_2( k_j , \Lambda) + \pi + \sum_{l=1}^N \Big[ \phi_1 \Big( k_j , k_l \Big) -  \phi_1 \Big(k_j^{(0)} , k_l^{(0)}\Big)\Big]. \label{Eq:difference}
\eea
Here the factor of $\pi$ arises from $\phi_2(k_j^{(0)},\Lambda_0)$ with $\Lambda_0 = -\infty$. 

We now use that the difference between the roots in the presence of the finite spin rapidity and those in the ground state are $k_j - k_j^{(0)} = O(L^{-1})$. We can then expand the scattering phase as 
\bea
\phi_1(k_j , k_l) = \phi_1(k_j^{(0)}, k_l^{(0)}) + K(k_j^{(0)},k_l^{(0)}) \Big[ \Big(k_j - k_j^{(0)}\Big) - \Big(k_l - k_l^{(0)}\Big)\Big],
\eea 
where we neglect terms $O(L^{-2})$ and the derivative of the phase is defined as 
\be
K(u,v) \equiv \phi'_1(u,v) = \frac{2c}{c^2 + (u-v)^2}.  
\ee
Keeping track of sums which pass to the continuum with symmetric (${\sum}'$) or non-symmetric ($\sum$) limits (see~\ref{App:limits}), and using 
\be
\sum_{l} \phi_1 \Big( k_j^{(0)} , k_l^{(0)}\Big) -  {\sum_{l}}' \phi_1 \Big(k_j ^{(0)}, k_l^{(0)}\Big) = O(L^{-1}),
\ee
we arrive at
\be
0 = \Big( k_j - k_j^{(0)} \Big) 2\pi L \rho\Big(k_j^{(0)}\Big) - \phi_2( k_j , \Lambda) + \pi - \sum_{l} \Big(k_l - k_l^{(0)}\Big)K\Big(k_j^{(0)},k_l^{(0)}\Big),
\label{Eq:difference2}
\ee
where we have kept terms to order $O(L^0)$ and we used the following identity for the root distribution 
\bea 
1 + \frac{1}{L} \sum_l K\Big(k_j^{(0)},k_l^{(0)}\Big) = 2\pi \rho\Big(k_j^{(0)}\Big). \label{Eq:root}
\eea
We now define the `shift function' (see, e.g., Ref.~\cite{KorepinBook}) 
\be
F_s\Big(k_j^{(0)}| \Lambda\Big) = \frac{k_j -  k_j^{(0)}}{k_j^{(0)}- k_{j-1}^{(0)}},
\ee
which can be interpreted physically as measuring the effect of the finite spin rapidity $\Lambda$ on the ground state momenta $k^{(0)}_j$. We pass to the continuum ($L\to\infty$ with $N/L$ fixed, see~\ref{App:limits} for details) 
and obtain the integral equation 
\bea
0 = 2\pi F_s\Big(k_j^{(0)}|\Lambda\Big) - \int^{q_R}_{-q_L} \rd q\, F_s(q|\Lambda) K\Big(k_j^{(0)},q\Big) -  \phi_2\Big(k_j^{(0)},\Lambda\Big) + \pi, \label{Eq:shifteq}
\eea
where we've used that the derivative term in the expansion of $\phi_2(k_j,\Lambda)$ about $k_j^{(0)}$ is $O(L^{-1})$, thus allowing us to drop it.

\subsubsection{The spin rapidity $\Lambda$.}
Let us recap how the spin rapidity $\Lambda$ is quantized. For a given integer $J$, the second Bethe equation~\fr{BA2} reads
\bea
2\pi J = \sum_{l=1}^N \phi_2 (\Lambda - k_l). 
\eea
Passing to the continuum according to Eqs.~\fr{Eq:continuum}, this becomes
\bea
2 \pi J = L \int_{-q_L}^{q_R} \rd k\, \rho(k) \phi_2(\Lambda,k) - \int_{-q_L}^{q_R} \rd k\, F_s(k|\Lambda) \phi_2'(\Lambda,k). \label{Eq:nu}
\eea
The second term is sub-leading in $L$ and we neglect it herein; this will be justified \textit{a posteriori} by direct comparison to the full numerical solution of the Bethe ansatz equations. 

\subsubsection{The dispersion relation.}

From the difference equation, it follows that the energy $E_s(\Lambda)$ and momentum $P_s(\Lambda)$ of the state (defined with respect to the ground state) with spin rapidity $\Lambda$ are given by
\be
E_s(\Lambda) = \int_{-q_L}^{q_R} \rd k\, 2 k F_s(k|\Lambda), \qquad P_s(\Lambda) = \int_{-q_L}^{q_R} \rd k\,  F_s(k|\Lambda), \label{Eq:EnuPnu}
\ee
where $F_s(k|\Lambda)$ satisfies the integral equation~\fr{Eq:shifteq}, with the root distribution $\rho(k)$ determined from Eq.~\fr{Eq:root} and $\Lambda$ the solution of Eq.~\fr{Eq:nu}. The Fermi momenta $q_L, q_R$ can be determined from the following relations:
\be
\frac{N}{L} = \int_{-q_L}^{q_R} \rd k\, \rho(k), \qquad \frac{P}{L} = \int_{-q_L}^{q_R} \rd k\, k \rho(k),  \label{Eq:NP}
\ee
where the momentum $P = - \pi(N+2 J)/L$ is given by Eq.~\fr{Eq:P} and implicitly depends on the spin rapidity $\Lambda$. 

We now compute the root distribution $\rho(k)$, the Fermi momenta $q_{R,L}$ and the shift function $F(k|\Lambda)$ under a $1/c$ expansion. We find 
\bea
\rho(k) = \frac{1}{2\pi} + \frac{1}{2 \pi^2 c}(q_R + q_L) + O\left( c^{-2} \right) \equiv \frac{1}{2\pi} + \frac{1}{\pi c}\varrho + O\left( c^{-2} \right) , \\
q_R =  \pi \varrho \Bigg(1 + \frac{2\varrho}{c} \Bigg) - \frac{\pi}{L} \left( \frac12 + \frac{J}{N} \right)\Bigg(1- \frac{4\varrho}{c}\Bigg) + O\left( c^{-2} \right), \label{Eq:qR} \\
q_L =  \pi \varrho \Bigg(1 + \frac{2\varrho}{c} \Bigg) + \frac{\pi}{L} \left( \frac12 + \frac{J}{N} \right)\Bigg(1- \frac{4\varrho}{c}\Bigg) + O\left( c^{-2} \right), \label{Eq:qL} \\
F_s(k|\Lambda) = \frac{1}{2\pi} \phi_2(k,\Lambda) - \frac12 + \frac{1}{2\pi^2} \left[ G\left(\frac{2(q_R - \Lambda)}{c} \right) - G\left(\frac{2(q_L + \Lambda)}{c}\right) \right]  - \frac{q_R + q_L}{2\pi c} + \ldots, \label{Eq:ShiftExp}
\eea
where $\varrho = N/L$ is the average particle density and we define the function $G(x) = x \arctan(x) - \frac12 \log(1 + x^2)$.

\begin{figure}
\begin{center}
\includegraphics[width = 0.48\textwidth]{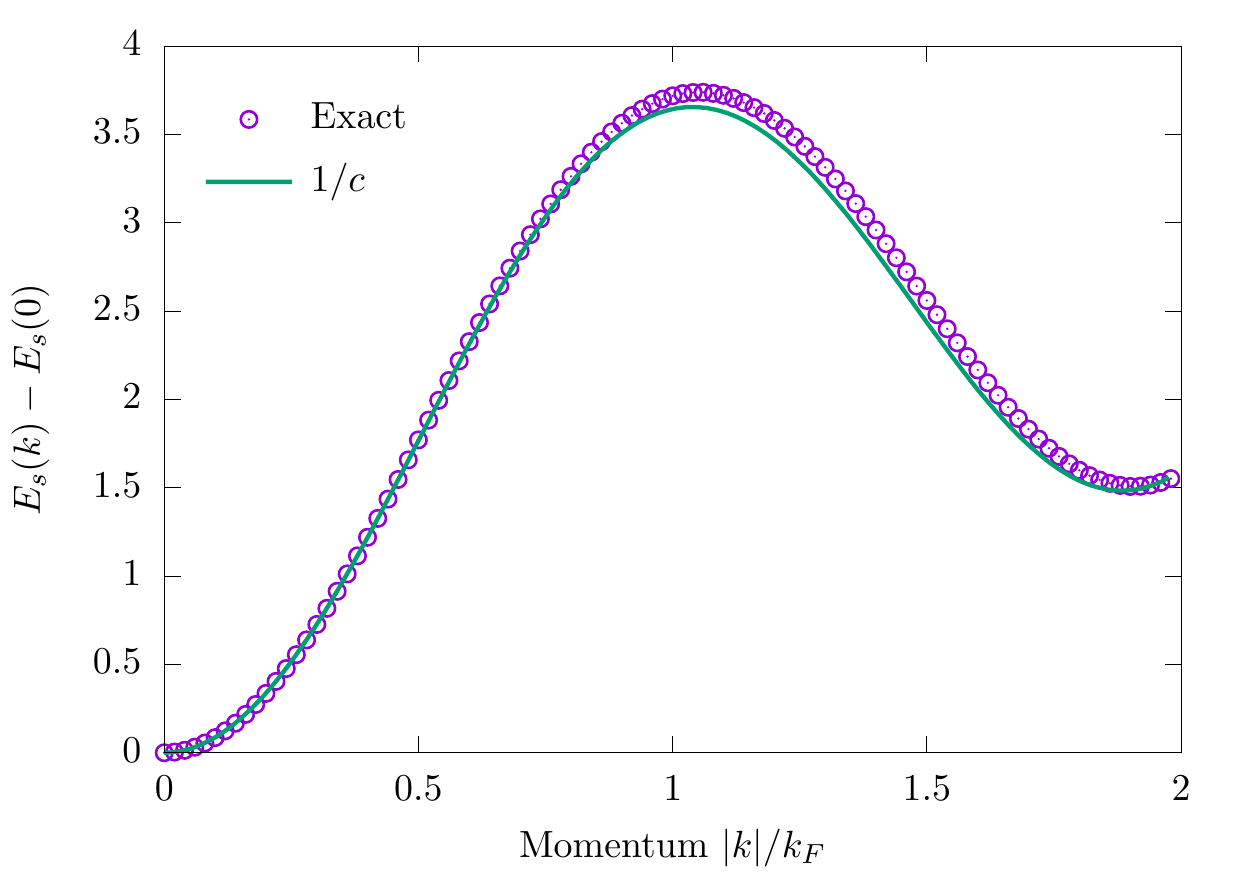}
\end{center}
\caption{(Line) The spin wave dispersion obtained from the strong coupling ($1/c$) expansion, computed from Eqs.~\fr{Eq:EnuPnu} using Eqs.~\fr{Eq:qR},~\fr{Eq:qL} and~\fr{Eq:ShiftExp}. (Points) The exact result for the spin wave dispersion obtained by direct numerical solution of the Bethe ansatz equations~\fr{BA1},~\fr{BA2} computed using Eqs.~\fr{Eq:P} and~\fr{Eq:E}. We have rescaled the momentum $k$ by the Fermi momentum $k_F = \pi \varrho$. Data is presented for $N=100$ particles on the length $L=50$ ring with interaction parameter $c=50$.}
\label{Fig:Comparison}
\end{figure}

Numerically integrating the set of equations~\fr{Eq:EnuPnu} containing the $1/c$ expansion for the shift function~\fr{Eq:ShiftExp} we obtain the energy and momentum for each state with spin rapidity $\Lambda$. We present the dispersion relation in Fig.~\ref{Fig:Comparison}, where we compare to the exact result for $N=100$ bosons on the length $L=50$ ring with interaction parameter $c=50$. The comparison shows that there is excellent agreement between the $1/c$ expansion computed above and the exact result. This validates our dropping of terms which are sub-leading in $L$ and $c$. 

At first glance, the spin wave dispersion is rather surprising. Let's first limit our attention to the well-studied region with $|k| \ll \varrho$. There we recognize the usual quadratic spin wave dispersion $E(k) \sim k^2/ 2m_*$ which can be understood from general symmetry considerations~\cite{HalperinPR69,HalperinPRB75}. The dependence of the effective mass $m_*$ on interaction strength is understood~\cite{FuchsPRL05,ZvonarevPRL07} -- at strong coupling, the effective mass diverges due to the `fermionization' (e.g., the hard core repulsion) of the bosons
\be
\frac{1}{2m_\ast} = \frac{1}{N} + \frac{4 \pi^2 \varrho}{3c} + O(c^{-2}).  \label{Eq:mstar}
\ee 
There is a simple picture for the mass $m_* \propto N$ being proportional to the number of particles $N$ in the strong coupling $c\to\infty$ limit. Consider a single flipped spin: due to the hardcore repulsion in the $c=\infty$ limit, one must move all the other particles on the ring in order to move the particle with different spin. As a result, a flipped spin acts much like a particle with mass $N$~\cite{FuchsPRL05}. 

Moving our attention away from the well-studied $|k| \ll \varrho$ region, the spin wave dispersion becomes non-monotonic with a gapped roton-like minima at close to $|k| \sim 2k_F$. The excitations about the roton-like minima have the dispersion
\be
E_r(p) = \Delta_r  + \frac{(p - p_r)^2}{2m_r}  + \ldots, \label{Eq:roton}
\ee
where $p_r$ is the momentum of the roton-like minima, $m_r$ is the effective mass for excitations about the minima, $\Delta_r$ is the energy gap (herein the roton gap, $\Delta_r \approx 1.5$ in Fig.~\ref{Fig:Comparison}), and the ellipses refer to terms higher order in $(p - p_r)$. Notice that the roton gap is \textit{many times larger than $\delta E_{YG}$, the finite volume excitation gap in the Yang-Gaudin Bose gas,}  
\be
\delta E_{YG} \sim  \Bigg(\frac{2\pi}{L}\Bigg)^2 \Bigg( \frac{1}{\varrho L} + \frac{4\pi^2\varrho}{3c}\Bigg), \label{Eq:LevelSpacingYG}
\ee
which follows from the strong coupling expansion of the spinon effective mass $m_*$~\cite{FuchsPRL05}, Eq.~\fr{Eq:mstar}. For the data presented in Fig.~\ref{Fig:Comparison} the finite volume excitation gap is $\delta E_{YG} \approx 0.0085$. 

\subsection{Finite-size effects: vanishing of the roton-like minima}

Now we turn our attention to how the spinon dispersion varies with the system size. As we have seen in the previous section, see Fig.~\ref{Fig:Comparison}, there is a pronounced finite-momentum roton-like minima in the dispersion at close to $k = -2k_F$. The problem that we have considered, $N$ particles with a single impurity $M=1$, is clearly susceptible to finite-size effects: the density of the impurity species is non-zero in the finite system, whilst it vanishes in the infinite volume limit $L\to\infty$. Accordingly, we may expect some changes in the dispersion with varying system size, although the usual assumption is that there changes will be small and only quantitative -- they arise as a result of the finite volume excitation gap -- and no qualitative features change. Here, we will show that this assumption is not valid -- there are large finite-size effects present in the spinon dispersion for relatively large $L\sim100$ systems and \textit{the infinite volume limit is qualitatively different to the finite volume}. 
\begin{figure}
\begin{center}
\includegraphics[width = 0.48\textwidth]{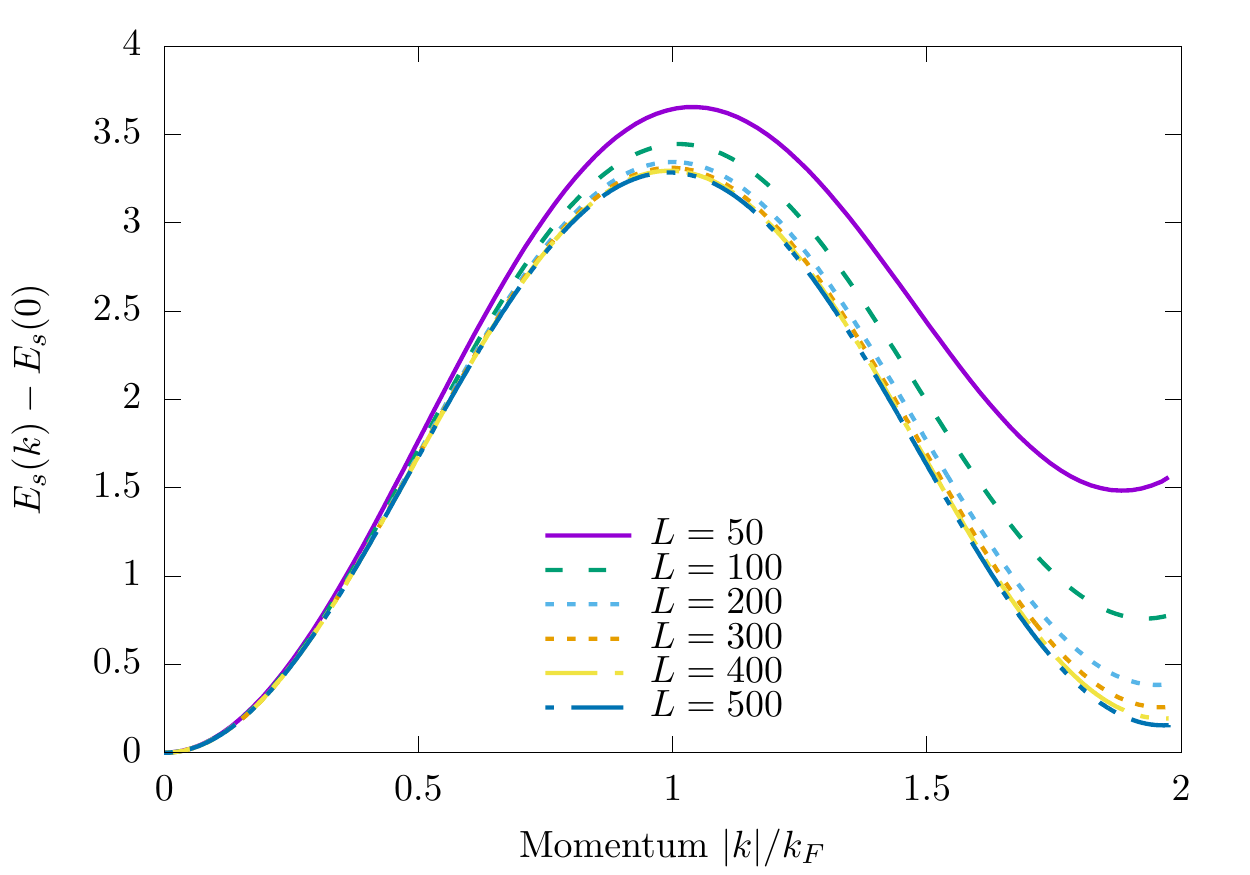}
\end{center}
\caption{The spin wave dispersion for $c=50$ from the $1/c$ expansion with fixed total particle density $\varrho \equiv N/L = 2$ at a number of system sizes $L=50 \to 500$. The roton-like minima shifts towards $-2k_F$ with increasing system size, whilst the gap vanishes as $1/L$.}
\label{Fig:Size}
\end{figure}

To begin, we present the spinon dispersion at fixed density $\varrho \equiv N/L = 2$ for a number of system sizes $L=50 \to 500$ in Fig.~\ref{Fig:Size}. We see that the roton-like minima shifts towards $2k_F$ with increasing system size, and the roton gap $\Delta_r$ is suppressed. In fact, the roton/$2k_F$ gap is suppressed as a power law in the system size: $\Delta_r \approx \Delta_{2k_F} \propto 1/L$. The value of the pre-factor can be computed directly from the strong coupling expansion (see the previous section) using that $k = -2k_F$ corresponds to $J = N/2$ and spin rapidity $\Lambda = \infty$. For $c$ large but finite, the shift function~\fr{Eq:ShiftExp} then becomes
\be
\lim_{\Lambda\to\infty} F_s(k|\Lambda) = - 1 - \frac{1}{2\pi c}(q_R+q_L),
\ee
and hence the energy~\fr{Eq:EnuPnu} at $k = -2k_F$ is
\be
\Delta_{2k_F} =  \frac{4\pi^2 \varrho}{L} \Bigg[ 1 + \frac{3\varrho}{c} + O\Big(c^{-2}\Big) \Bigg]. \label{Delta2kF}
\ee
Here we see both the scaling of the gap $\Delta_{2k_F} \sim 1/L$ and the origin of the large finite-size effects: the pre-factor $4\pi^2 \varrho \gg 1$ is large. The large pre-factor arises from the term $(q_R^2 - q_L^2)$ when the Fermi sea is imbalanced $q_R \neq q_L$, see Eqs.~\fr{Eq:qR}~and~\fr{Eq:qL}. Notice that there is not a strong $c$-dependence of the pre-factor; this is consistent with the roton/$2k_F$ gap being present for weak interactions, as seen in small systems in Ref.~\cite{LiEPL03}. 

\subsection{Roton-like excitations in the thermodynamic limit}
We have seen that the $2k_F$ gap (and the roton-like minima) vanish in the limit $L\to\infty$ when the spinon number $M=1$ is fixed. A natural question to ask is whether these features still vanish in the thermodynamic limit with fixed spinon density ($L\to\infty$ with $N/L,\ M/L$ fixed)? Physically, such a scenario corresponds to the case with finite population imbalance. In this section, we attempt to address this question. First, we will present a non-rigorous argument which suggests that, indeed, the roton-like excitations and $2k_F$ gap should persist in the thermodynamic limit with fixed spinon density. We follow this with supporting numerical data and a discussion of the physical origin of the roton-like minima. 

\subsubsection{A non-rigorous argument.} 
Here we will show that the Bethe ansatz equations for two almost-identical spinons approximately reduce to twice the Bethe ansatz equations for a single spinon when working at fixed density $N/L, M/L$. Let us begin by writing the single spinon Bethe ansatz equations, defined by the integers $I^{(1)}_j$ and $J^{(1)}$ for $N$ particles with $M=1$ on the length $L$ ring: 
\bea
2\pi I^{(1)}_j &=& k^{(1)}_j L + \sum_{l=1}^N \phi_1 \Big(k_j^{(1)},k_l^{(1)} \Big) - \phi_2 \Big(k_j^{(1)},\Lambda^{(1)} \Big), \\
2 \pi J^{(1)} &=& \sum_{l=1}^N \phi_2 \Big(\Lambda^{(1)},k_l^{(1)} \Big). \label{Eq:spin1}
\eea

We next consider the case with double the number of spinons, $M=2$. We want to consider two close-to-identical spinons, so we choose neighboring values for the spin rapidity integers: $J^{(2)}_1 = J_2^{(2)}-1$. As we have doubled the number of spinons, we must also double $L$ and $N$ to continue working at fixed particle and spinon density. Recall that this means the range of the integers $I_j^{(2)}$ and $J^{(2)}$ is also doubled in comparison to the case with $M=1$. The Bethe ansatz equations in this case read
\bea
2 \pi I^{(2)}_j &=& 2 k^{(2)}_j L + \sum_{l=1}^{2N} \phi_1 \Big(k_j^{(2)},k_l^{(2)} \Big) - \sum_{\beta=1}^2 \phi_2 \Big(k_j^{(2)},\Lambda^{(2)}_\beta \Big), \label{Eq:I2} \\
2\pi J^{(2)}_\alpha &=& \sum_{l=1}^{2N} \phi_2 \Big(\Lambda^{(2)}_\alpha,k_l^{(2)} \Big) - \sum_{\beta =1}^2 \phi_1 \Big(\Lambda_\alpha^{(2)},\Lambda_\beta^{(2)} \Big), ~~~\alpha =1,2. \label{Eq:J2}
\eea
There are now a number of points to note. Firstly, taking the sum and difference of the lower equations, we have 
\bea
2\pi \Big( 2 J^{(2)}_1 + 1 \Big) = \sum_{l=1}^{2N} \sum_{\alpha = 1}^2 \phi_2 \Big(\Lambda^{(2)}_\alpha,k_l^{(2)} \Big), \label{Eq:sum2} \\
2\pi = \sum_{l=1}^{2N} \sum_{\alpha=1}^2 (-1)^\alpha \phi_2 \Big(\Lambda^{(2)}_\alpha,k_l^{(2)} \Big) - 2 \phi_1 \Big(\Lambda_2^{(2)},\Lambda_1^{(2)} \Big). \label{diff}
\eea
Taking the continuum limit, the first term on the right hand side of Eq.~\fr{diff} acquires a factor of $2L$ [cf. Eq.~\fr{Eq:nu}] which implies that $\Lambda_2^{(2)} - \Lambda_1^{(2)} = O(L^{-1})$. Accordingly, we parameterize the spin rapidities by 
\be
\Lambda_\alpha^{(2)} = \Lambda^{(2)} +( -1)^\alpha d \Lambda \quad \mathrm{with}\quad d\Lambda \sim O(L^{-1}).
\ee
We can now expand the right hand side of Eq.~\fr{Eq:sum2} to give 
\be
2\pi \Big( 2J_1^{(2)} + 1 \Big) = 2 \sum_{l=1}^{2N} \phi_2 \Big(\Lambda^{(2)},k_l^{(2)}  \Big) + O\Big( (d\Lambda)^2 \Big).
\ee
We can simplify this equation in two further manners. Firstly, as we are working at fixed particle density $N/L$, the Fermi surface for the momenta quantum numbers is unchanged (up to corrections of order $1/L$) and hence increasing the particle number simply increases the density of the momenta within the Fermi surface. We can approximate to leading order the sum as 
\be
\sum_{l=1}^{2N} \phi_2 \Big(\Lambda^{(2)},k_l^{(2)} \Big) \approx 2 \sum_{l=1}^N \phi_2 \Big(\Lambda^{(2)},k_l^{(1)} \Big). 
\ee
Secondly, we can approximate $J_1^{(2)}$ by $2J^{(1)}$ up to an additive factor of unity. As the right hand side will be proportional to $L$ in the continuum limit, single factors of one are unimportant. Thus, we can approximate Eq.~\fr{Eq:sum2} by
\be
2 \pi J^{(1)} \approx \sum_{l=1}^N \phi_2 \Big(\Lambda^{(2)},k_l^{(1)} \Big),
\ee
which is the same equation as for the spin rapidity in the $M=1$ case, see Eq.~\fr{Eq:spin1}. In other words, the spinon rapidity is (approximately) the same for the case with $M=1$ and $M=2$ provided the spinon density is identical. Working through the same arguments for Eq.~\fr{Eq:I2}, we find that the $k_{2j}^{(2)}, k_{2j+1}^{(2)} \approx k_j^{(1)}$ and hence the energy and momentum of a two spinon states is approximately described by $E^{(2)} \approx 2 E^{(1)}$, $P^{(2)} \approx 2 P^{(1)}$, as would naively be expected for two almost identical spinons. This gives some support to the idea that the roton gap for the single spinon excitation can persist to infinite volume provided both $N/L$ and $M/L$ are fixed.

\begin{figure}
\begin{center}
\includegraphics[width=0.48\textwidth]{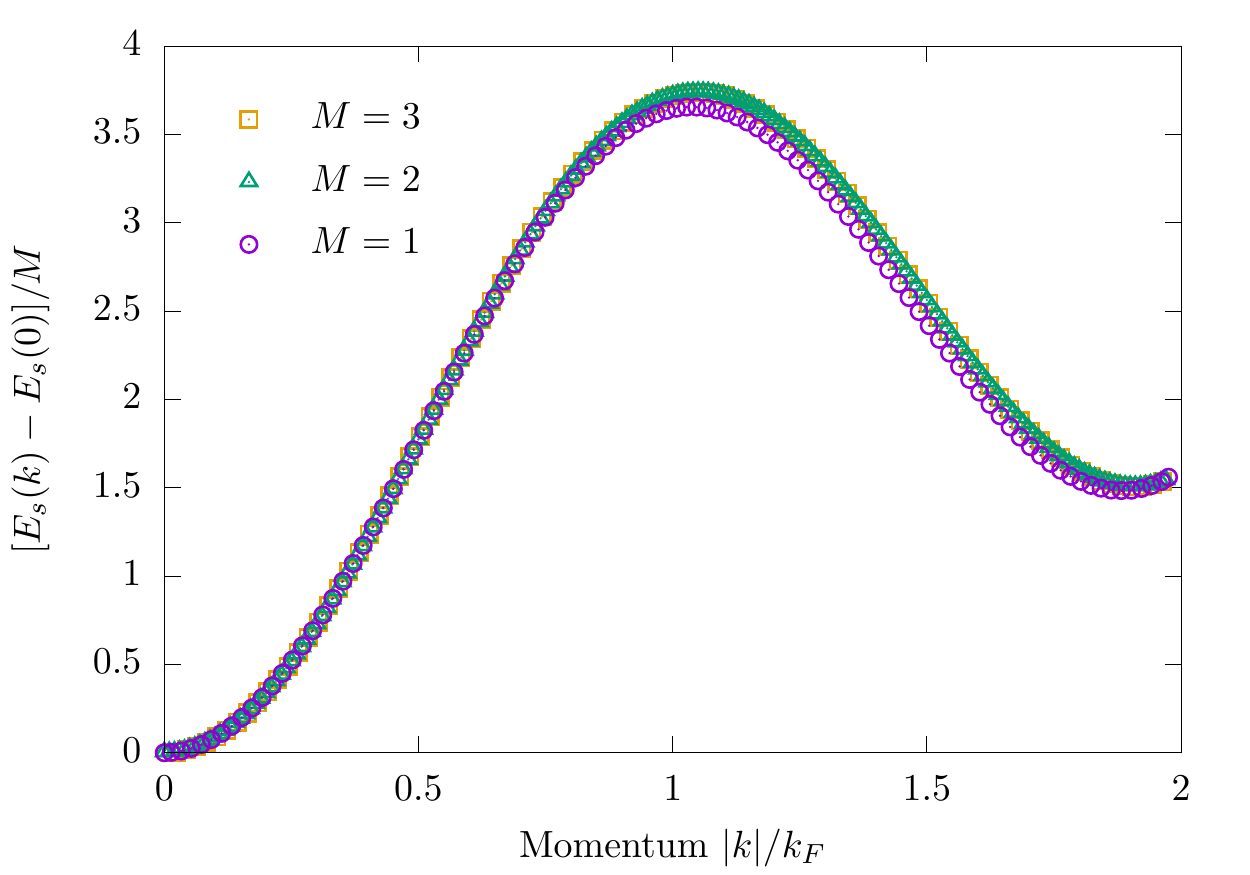}
\end{center}
\caption{Comparison of the spinon dispersion extracted from the full solution of the Bethe ansatz equations with $M=1$, $M=2$ and $M=3$ at fixed particle density $N/L = 2$ and fixed spinon density $M/L = 1/50$. We compute the dispersion relation for $M$ ``almost identical'' spinons by setting the spin rapidity quantum numbers to be neighboring, e.g., for $M=2$ we choose $J_1 = J_2-1$ and for $M=3$ we choose $J_1 = J_2 -1 = J_3-2$, cf. Eqs.~\fr{FullBA1}~and~\fr{FullBA2}.}
\label{Fig:MultiSpinon}
\end{figure}

\subsubsection{Numerical supporting evidence.} 
The above argument, whilst suggestive that the roton gap persists to infinite volume, is not rigorous. To confirm that this argument is essentially valid, we provide supporting evidence from the numerical exact solution of the full Bethe ansatz equations~\fr{FullBA1},~\fr{FullBA2} for the case of $M=2,3$ ``almost identical'' spinons (that is, we choose the spin rapidity quantum numbers to be $J_1 = J_2-1$ for $M=2$ and $J_1=J_2-1=J_3-2$ for $M=3$) at constant particle density $N/L$ and spinon density $M/L$. In Fig.~\ref{Fig:MultiSpinon} we present the extracted single spinon dispersion from computation of the energy and momentum of the ``almost identical'' spinon states. We see that there is an approximate collapse of results upon rescaling by the spinon number $M$, when working at fixed particle and spinon density. This supports the idea that the roton-like minima and $2k_F$ gap persist to the thermodynamic limit when the spinon density is finite (while maintaining the limit $M \ll N$).\footnote In Sec.~\ref{Sec:YYT}, we will see that finite temperature in the thermodynamic limit also opens a $2k_F$ gap in the dressed spinon energy. The spin wave excitations in the Yang-Gaudin Bose gas at fixed spinon density $M/L$ thus constitute an example of roton-like excitations  in an exactly solvable microscopic many-body model, a problem which has attracted attention for over 75 years~\cite{LandauZETF41}--\cite{CohenPR57}. 

\subsubsection{Origin of the roton-like minima.}
\label{Sec:Origin}
Having established the presence of a $2k_F$ gap and roton-like minima in the spin wave dispersion, it is natural to ask what is their physical origin? At a technical level, the behavior is quite easy to explain -- everything can be understood in terms of the integers appearing in the Bethe equations, Eqs.~\fr{BA1}~and~\fr{BA2}. To begin, it is useful to consider the ground state with $J_0 = -N/2$, corresponding to a spin rapidity of $\Lambda = -\infty$. Then, the Bethe equation~\fr{BA1} becomes
\be
2\pi \Big(I_j^{(0)} + \frac12\Big) \equiv 2\pi \tilde I_j^{(0)} = k_j^{(0)} L + \sum_{l=1}^N \phi_1(k_j,k_l), \label{GSBA1}
\ee
where we define the shifted `integers' $\tilde I_j^{(0)} = I_j^{(0)} + 1/2$. The ground state is realized as the symmetric Fermi sea of the shifted integers $\tilde I^{(0)} = \left\{ -\frac{N-1}{2}, \ldots, \frac{N-1}{2} \right\}$, and the solution of Eq.~\fr{GSBA1} is in one-to-one equivalence with the ground state of the (one-component) Lieb-Liniger model~\cite{LiebPR63a}--\cite{KorepinBook}. Now, let us turn our attention to the gap at $k\sim-2k_F$, which corresponds to solutions with large spin rapidity $\Lambda \gg 1$. As a result, the Bethe equation~\fr{BA1} can be approximated by
\be
2\pi \Big(I_j^{(0)} - \frac12\Big) \equiv 2\pi \Big(\tilde I_j^{(0)}-1\Big) \approx k_j^{(0)} L + \sum_{l=1}^N \phi_1(k_j,k_l). \label{ShiftedBA1}
\ee
The solutions to Eq.~\fr{ShiftedBA1} are formally equivalent to the scenario where the right-most shifted integer of the ground state configuration $\tilde I^{(0)}$ is scattered across the Fermi sea: $\tilde I^{(0)}_N = \frac{N-1}{2} \to -\frac{N+1}{2}$. The momentum $-2k_F$ and energy gap $\Delta_{2k_F}$ thus coincide with that of the holon-antiholon $-2k_F$ excitation in the (one-component) Lieb-Liniger model (see also the following section)~\cite{LiebPR63a}--\cite{KorepinBook}.

\begin{figure}
\begin{center}
\begin{tabular}{ll}
\includegraphics[width=0.48\textwidth]{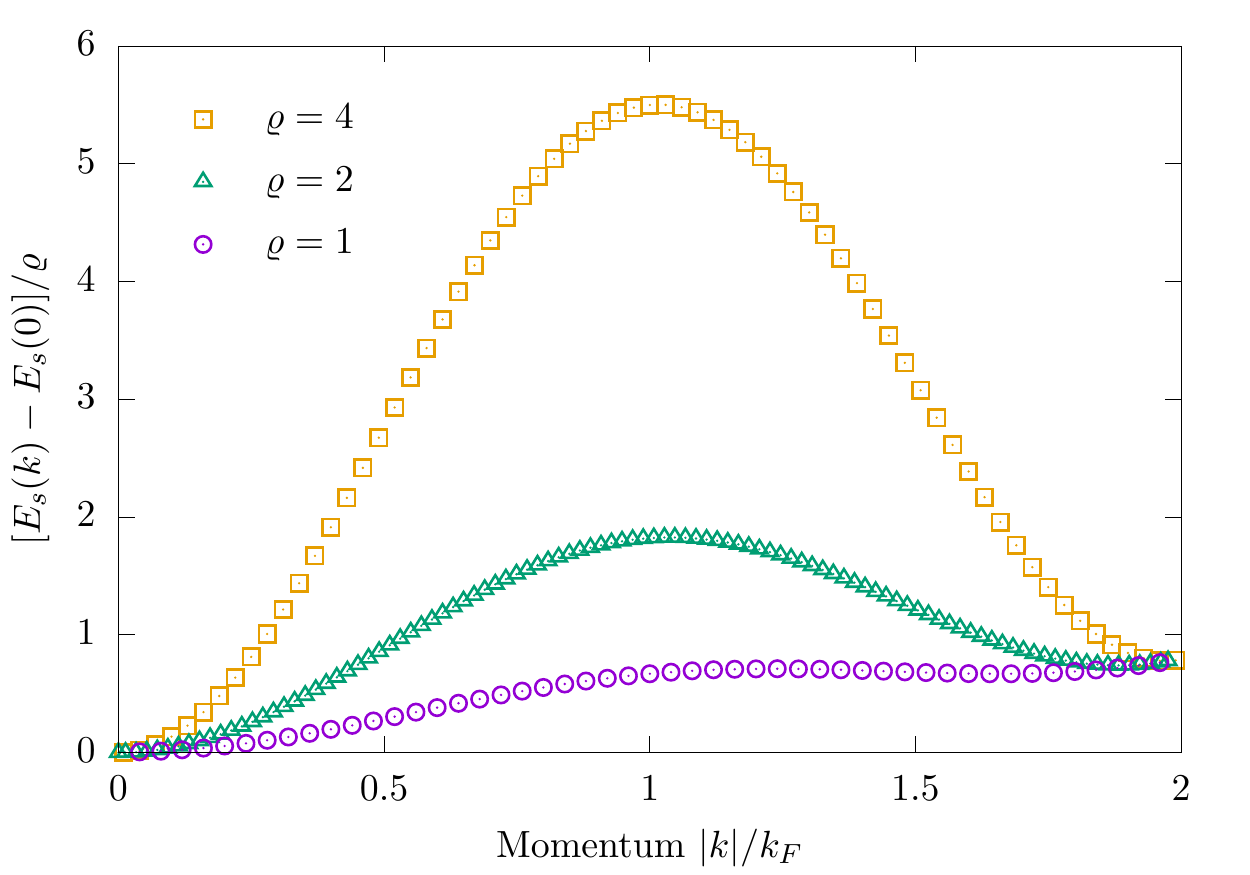} &
\includegraphics[width=0.48\textwidth]{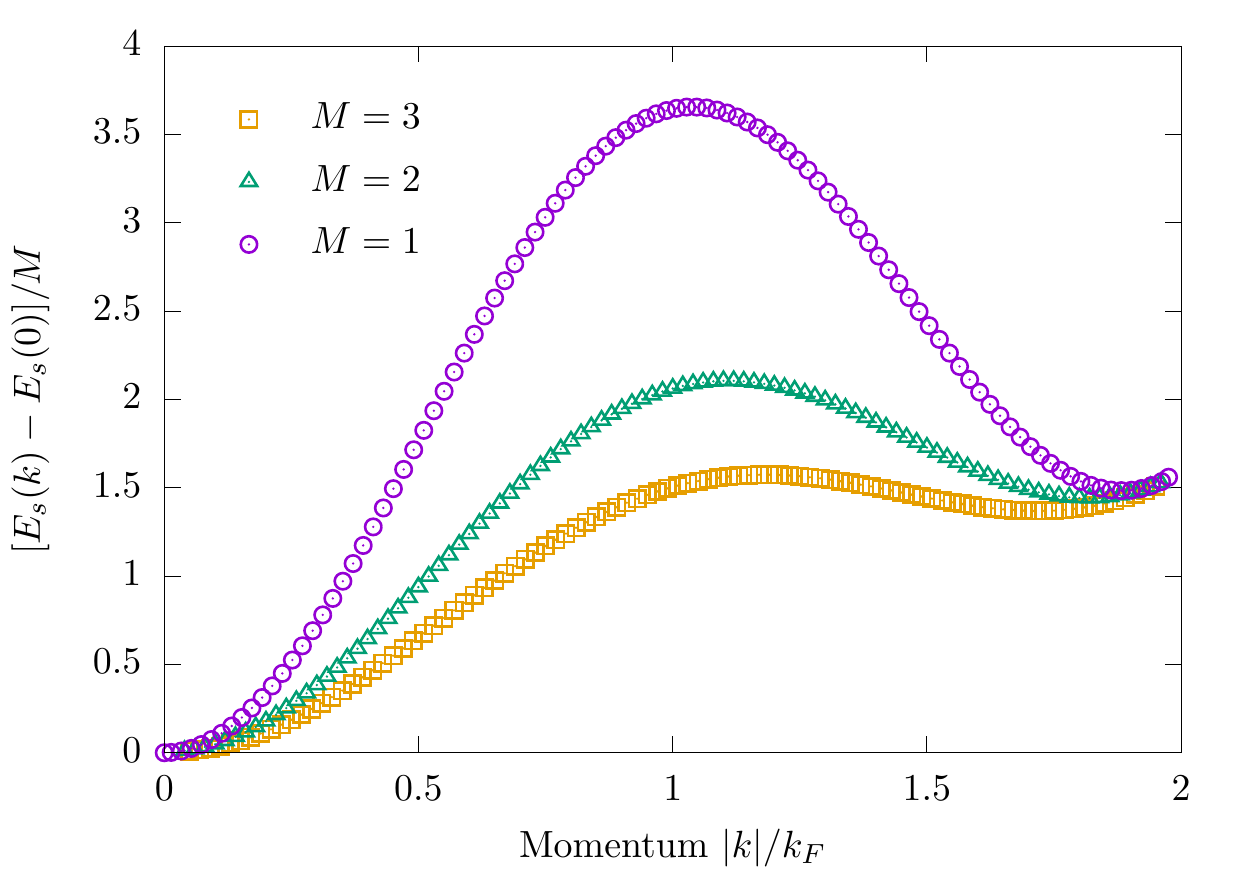}
\end{tabular}
\end{center}
\caption{(Left panel) Scaling of the spinon dispersion $E_s(k)$ with varied particle number density $\varrho = N/L$ for $M=1$, $L=c=50$.  (Right panel) Scaling of the spinon dispersion $E_s(k)$ with varied spinon density $\varrho_s = M/L$ for $N=100$, $L=c=50$. We see that the $2k_F$ gap in the single spinon dispersion relation varies linearly with the density of background gas $\varrho$ and the spinon density $\varrho_s$, both consistent with the strong coupling expansion~\fr{Delta2kF} for $M=1$. }
\label{Fig:VaryDensity}
\end{figure}

At a formal level we see that the $2k_F$ gap can be pictured in a similar manner to the $-2k_F$ holon-antiholon excitation of the (one-component) Lieb-Liniger model, but a more physical explanation would be nice. In the above, we clearly see that presence of interactions between the constituent excitations of the Yang-Gaudin Bose gas (e.g., spinons and (anti)holons) plays an important role. The role of interactions can be further elucidated by considering how the spinon dispersion varies with density of spinons, the number of particles, and the system size. To begin, we fix the spinon density $\varrho_s = 1/L$ and vary the total particle density $\varrho$. We see in Fig.~\ref{Fig:VaryDensity} that the $2k_F$ gap $\Delta_{2k_F}$ is proportional to the total particle density $\varrho$, as can be expected from the strong coupling expansion~\fr{Delta2kF}. If instead we fix the total particle density and increase the spinon density~$\varrho_s$ (by consider the ``almost identical'' spinon configuration discussed in the previous sections), we find similar: $\Delta_{2k_F}$ is proportional to the spinon density. This is also consistent with the strong coupling expansion for $M=1$ ($\varrho_s = 1/L$), where $\varrho_s$ enters through the imbalance of the Fermi points, $q_R-q_L$. Combining these results, we see that the $2k_F$ gap is proportional to the product of the spinon and particle densities, consistent with the gap being induced by spinon-(anti)holon interactions.

\subsection{Finite-size effects: the finite volume excitation gap in the Yang-Gaudin Bose gas}
Before moving on to discuss two particle excitations, it is useful to briefly discuss the finite volume excitation gap in the Yang-Gaudin model and compare to that in the (one-component) Lieb-Liniger model. Henceforth, when we say the  \textit{finite volume excitation gap} we mean the minimum excitation energy above the ground state (for any type of excitation) when there is a fixed number of particles $N$.\footnote{The name reflects the fact that excitations are gapless in the infinite volume limit with fixed density.} 

In the Yang-Gaudin model at strong coupling, the finite volume excitation gap is set by the spinon excitation and can be derived directly from the strong coupling expression for the effective spinon mass $m_\ast$~\cite{FuchsPRL05} (see Eq.~\fr{Eq:mstar})
\be
\delta E_{YG} \sim \left(\frac{2\pi}{L}\right)^2 \left(\frac{1}{\varrho L} + \frac{4 \pi^2 \varrho}{3c}\right) + O(c^{-2}),
\ee
see also Eq.~\fr{Eq:LevelSpacingYG} and the accompanying discussion. 

On the other hand, in the (one-component) Lieb-Liniger model there exist only holon/antiholon excitations. Here, the finite volume excitation gap corresponds to taking the momentum integer at the Fermi surface $I_{FS}$ and moving it to $I_{FS}+1$ (creating, effectively, a $h\bar h$ excitation, see the following section). As a result, the finite volume excitation gap of the (one-component) Lieb-Liniger model is 
\be
\delta E_{LL} \sim \left(\frac{2\pi}{L}\right)^2 \left(1 - \frac{4\varrho}{c}\right) \varrho L  + O(c^{-2}), \label{Eq:ELL}
\ee
as can be found from, e.g., Eq.~\fr{Eq:Ehhbar}. We note that the Yang-Gaudin model with $M=0$ reduces to the one-component Lieb-Liniger model and as a result, the holon-antiholon minimum excitation energy in the Yang-Gaudin model is also given by Eq.~\fr{Eq:ELL}.

Let us now briefly highlight an important point. Setting $c = \infty$ for clarity, we have 
\be
\frac{\delta E_{YG}\big|_{c=\infty}}{\delta E_{LL}\big|_{c=\infty}} = \frac{1}{N^2}. \label{Eq:LevelSpacingRatio}
\ee
That is, in the hard core limit the finite volume excitation gap of the Yang-Gaudin model (set by the spinon excitation) is always much smaller than that of the (one-component) Lieb-Liniger model.\footnote{In the Yang-Gaudin Bose gas at strong coupling, Eq.~\fr{Eq:LevelSpacingRatio} simply states that the finite volume excitation gap for the spinon excitation is always much smaller than the minimal excitation energy for the holon-antiholon excitation.} In other words, there will be many low-energy spinon states below the energy scale $\delta E_{LL}$ in the Yang-Gaudin Bose gas. 

\subsection{Extending beyond $2k_F$}

We have considered the dispersion of excitations described by the configuration of integers~\fr{Eq:Is}, in which the momenta integers $\{I\}$ are described by the ground state configuration and the integer $J_s$ is varied between its bounds $|J_s|\le N/2$. With fixed momenta integers $\{I\}$, the bounding of $J_s$ implies the spinon excitation has bounded momentum $0 \le |k| \le 2k_F$. To realize \textit{spinon-like} states with higher momenta, it is necessary to modify the configuration of the momenta integers $\{I\}$, creating multi-particle excitations. To extend to the momentum range $2k_F \le |k| \le 4k_F$, we remove ($-$) the momenta integer at the right Fermi point and add ($+$) it immediately to the left of the left Fermi point, creating a three-particle $sh\bar h$ excitation. For $N=8$ particles, such a configuration of integers is shown below:  
\begin{center}
        \integers
        \border
        \intrange{-9}{8}
        \fillI{-4}{3}
        \holeadd{3}{-}
        \particleadd{-5}{+}
        \intLabel{I}
        \render
\end{center}
\begin{center}
        \integers
        \intrange{-4}{4}
        \fillI{-2}{-2}
        \particleadd{-2}{J_s}
        \intLabel{J}
        \render
\end{center}

\begin{figure}
\begin{center}
\includegraphics[width=0.48\textwidth]{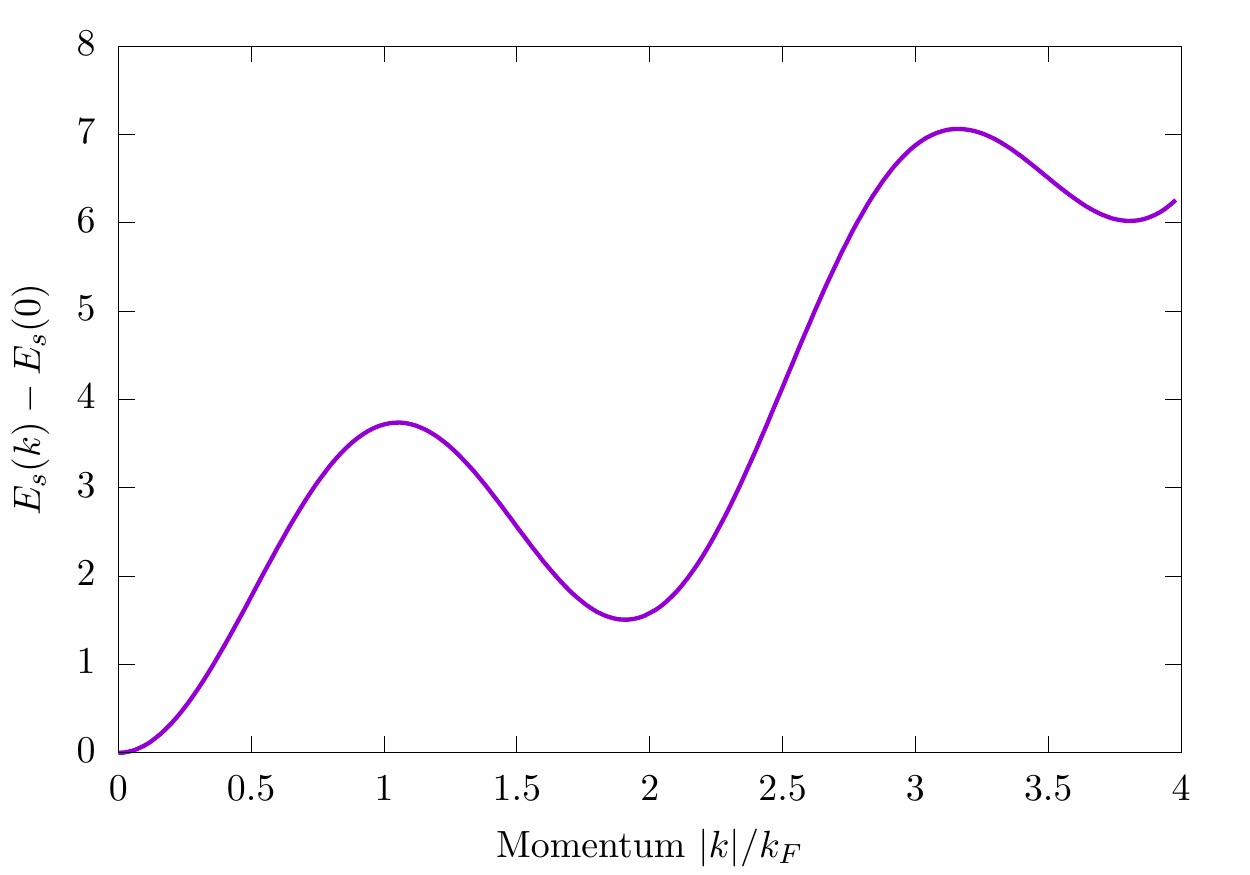}
\end{center}
\caption{The dispersion relation for spinon and spinon-like states computed by the numerical solution of the Bethe ansatz equations~\fr{BA1},~\fr{BA2} (cf. Fig.~\ref{Fig:Comparison} for the region $0\le|k|\le 2k_F$). We compute the dispersion for $N=100$ bosons on the length $L=50$ ring with interaction strength $c=50$.}
\label{Fig:Extended}
\end{figure}

In Fig.~\ref{Fig:Extended} we present the dispersion relation for the spinon ($0\le|k|\le2k_F$) and the spinon-like ($2k_F < |k|\le 4k_F$) states computed from the numerical solution of the Bethe ansatz equations~\fr{BA1},~\fr{BA2} for $N=100$ bosons on the length $L=50$ ring with interaction parameter $c=50$. We see that the roton-like minima in the spinon dispersion is indeed a local minima, with the dispersion of the spinon-like states increasing, before the appearance of an additional roton-like minima close to $4k_F$. Computing the dispersion for spinon-like states carrying higher momentum (generated by translating the Fermi sea of integers further to the left), we see that the spinon-like states have dispersion $\sim k^2$ with additional slowly-decaying oscillations superimposed on top of this trend. 

We note that the spinon-like excitations considered here are absolutely stable, having an infinite lifetime due to the integrability of the model. This prevents the decay of a spinon-like excitation with energy $E>2\Delta_r$ into two roton-like excitations, an instability that occurs in non-integrable models (and is observed experimentally in high-density liquid Helium~\cite{DonnellyJLowTempPhys81}) that results in the appearance of a Pitaevskii plateau~\cite{PitaevskiiSovPhysJETP59} in the dynamical structure factor. Related phenomena, known as \textit{quasi-particle breakdown}, is also observed in spin systems when one- and two-particle excitations overlap~\cite{ZhitomirskyRMP13}.

\section{Two particle excitations}
\label{Sec:TwoParticle}

Having observed large finite-size effects in the single particle dispersion, we now turn our attention to computation of the two-particle excitation continuum for the $h\bar h$ and $s h$ excitations. In particular, we will focus on whether there exists a finite energy gap at $|k| = 2k_F$ in the thermodynamic limit.

\subsection{The holon-antiholon ($h \bar h$) continuum}

The $h \bar h$ continuum constructed on top of the absolute ground state~\fr{Eq:IntegersGS} is identical to the particle-hole excitation continuum of the Lieb-Liniger model (see, e.g., Ref.~\cite{KorepinBook}). Notice also that setting the spin rapidity $\Lambda = -\infty$, we recover the Bethe ansatz equations for the Lieb-Liniger model and  so the $h\bar h$ continuum above this state is also equivalent to that in the Lieb-Liniger model. We consider the problem of removing a particle with momentum $-q_0 < k_h < q_0$ from within the Fermi sea and replacing it with a particle outside the Fermi sea $|k_p| > q_0$. The two-particle continuum of excitations is characterized in terms of the energy $E(k_p,k_h)$ and momentum $P(k_p,k_h)$ given by~\cite{KorepinBook}
\bea
E_{h\bar h}(k_p,k_h) = k_p^2 - k_h^2 - \int_{-q_0}^{q_0} \rd k\, 2 k\ F_{h\bar h}(k|k_p,k_h), \label{Eq:Eph}\\
P_{h\bar h}(k_p,k_h) = k_p - k_h - \int_{-q_0}^{q_0} \rd k\, F_{h\bar h}(k|k_p,k_h). \label{Eq:Pph}
\eea
Here we define $F_{h\bar h}(k|k_p,k_h) = (k_j - \bar k_j)/(k_j - k_{j-1})$ the shift function defined with ground state momenta $k_j$ and the excited state (with the $h\bar h$ excitation) momenta $\bar k_j$. The shift function obeys the following integral equation~\cite{KorepinBook}
\bea
F_{h\bar h}(q|k_p,k_h) - \int_{-q_0}^{q_0} \frac{ \rd k}{2\pi}\, K(q,k)F_{h\bar h}(k|k_p,k_h)= \frac{1}{2\pi} \Big[ \phi_1(q-k_p) - \phi_1(q-k_h) \Big].
\eea

\subsubsection{Strong coupling expansion.}
We can now compute the $1/c$ expansion for the relevant quantities, as done when considering the spinon dispersion. We find
\bea
\rho(k) = \frac{1}{2\pi} \Bigg(1 + \frac{2\varrho}{c}\Bigg)+ O(c^{-3}), \\
q_0 = \pi \varrho \Bigg(1 - \frac{2\varrho}{c} + \frac{4\varrho^2}{c^2}\Bigg)+ O(c^{-3}), \label{Eq:q0ph}\\
F_{h\bar h}(k|k_p,k_h) = \frac{1}{\pi c} (k_h-k_p) \Bigg[ 1 + \frac{2\varrho}{c} \Bigg] + O(c^{-3}).
\eea
For a configuration of integers $\{I_1,\ldots, I_N \}$, the strong coupling expansion for the momenta is
\bea
k_j = \frac{2\pi I_j}{L} \Bigg( 1- \frac{2\varrho}{c} + \frac{4\varrho^2}{c^2} \Bigg) + O(c^{-3}),
\eea
which is consistent with our expression for the Fermi momentum $q_0$, Eq.~\fr{Eq:q0ph}.
\begin{figure}
\begin{center}
\includegraphics[width=0.48\textwidth]{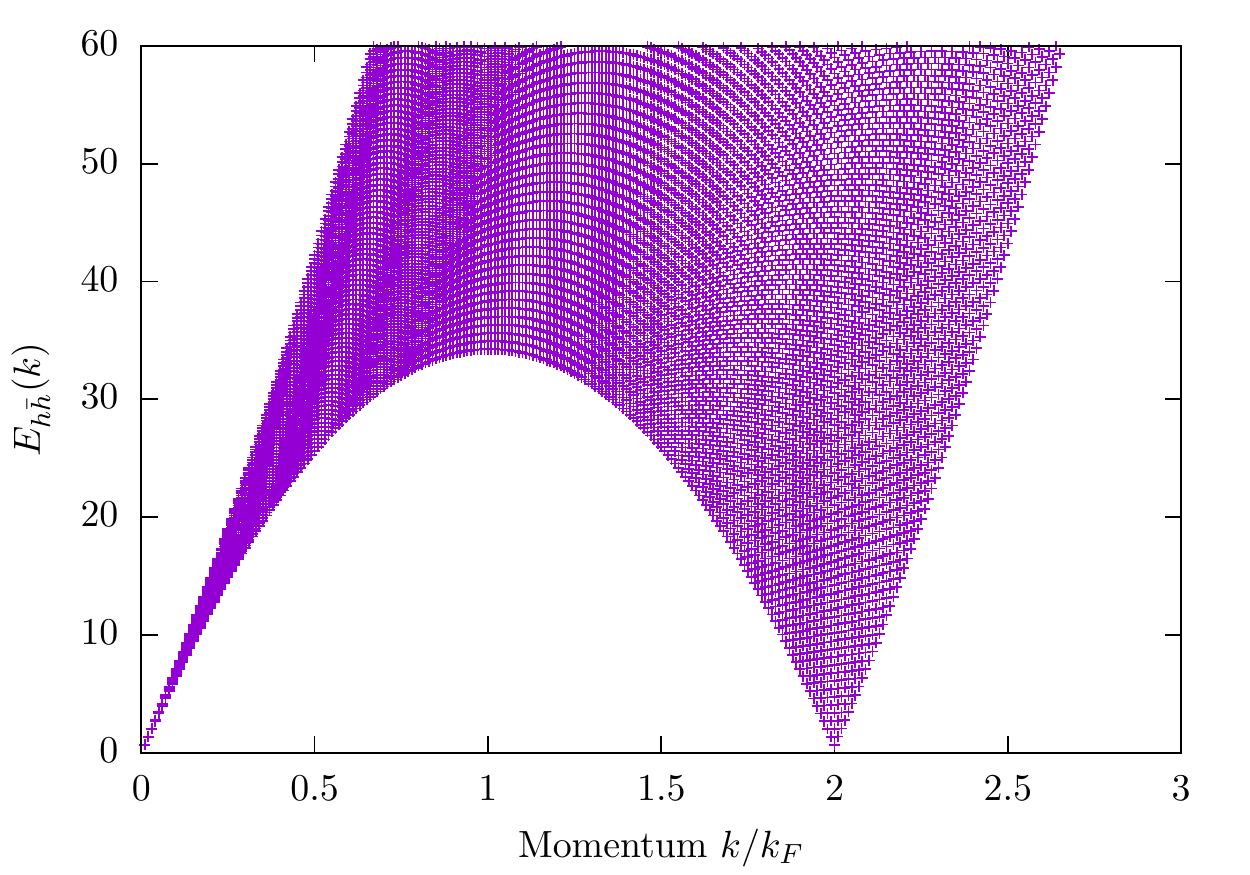}
\end{center}
\caption{The $h\bar h$ two-particle excitation continuum for system size $L=100$ with $\varrho \equiv N/L =2$ and interaction strength $c=50$. }
\label{Fig:ph}
\end{figure}

As the shift function is functionally independent of $k$ to order $c^{-2}$, we can directly integrate Eqs.~\fr{Eq:Eph} and~\fr{Eq:Pph} to obtain
\be
E_{h\bar h}(k_p,k_h) = k_p^2 - k_h^2 + O(c^{-3}),\qquad P_{h\bar h}(k_p,k_h) = \Big( k_p - k_h \Big) \Bigg[ 1 + \frac{2\varrho}{c}  \Bigg] + O(c^{-3}).
\ee
Denoting the hole in the Fermi sea of integers by $-\frac{N-1}{2} \le I_h \le \frac{N-1}{2}$ and the excited integer by $I_p > \frac{N-1}{2}$, we have the simple relations
\bea
E_{h\bar h}(I_p,I_h) = \Bigg(\frac{2\pi}{L}\Bigg)^2 \Bigg(1 - \frac{4\varrho}{c} + \frac{12 \varrho^2}{c^2} \Bigg) (I_p^2 - I_h^2 ) + O(c^{-3}), \label{Eq:Ehhbar}\\
P_{h\bar h}(I_p,I_h) = \frac{2\pi}{L} (I_p - I_h) + O(c^{-3}).
\eea
This realises the usual holon-antiholon continuum of the Lieb-Liniger model~\cite{KorepinBook}, shown in Fig.~\ref{Fig:ph}. From Eq.~\fr{Eq:Ehhbar}, the sole effect of finite interaction strength [to order $O(c^{-3})$] is to rescale the energy compared to the $c=\infty$ limit; for $\varrho=2$, $c=50$ (as shown in Fig.~\ref{Fig:ph}) the energy is rescaled by $(1-2\varrho/c + 4\varrho^2/c^2) \approx 0.85$, e.g., $15\%$. As is well known in the Lieb-Liniger model~\cite{KorepinBook}, there is a gap at $k=2k_F$ due to the finite volume excitation gap -- the lowest energy $2k_F$ excitation is umklapp-like: an integer is removed at the left Fermi point $-k_F$ and inserted immediately to the right of the Fermi point at $k_F$. The state is thus separated from the ground state by the finite volume excitation gap $\delta E_{LL}$ given in Eq.~\fr{Eq:ELL} (cf. Eq.~\fr{Eq:LevelSpacingYG}). With increasing system size $L$, there is no qualitative change in the $h\bar h$ continuum, unlike in the spinon dispersion (cf. Fig.~\ref{Fig:Size}). We note that the $2k_F$ gap in the $h\bar h$ continuum is comparable in size to the roton gap for the spinon excitations $\delta E_{LL} \approx \Delta_r$, cf. Eq.~\fr{Eq:ELL} and Eq.~\fr{Delta2kF} 

\subsection{The spinon-holon ($sh$) continuum}

\begin{figure}
\begin{center}
\includegraphics[width=0.48\textwidth]{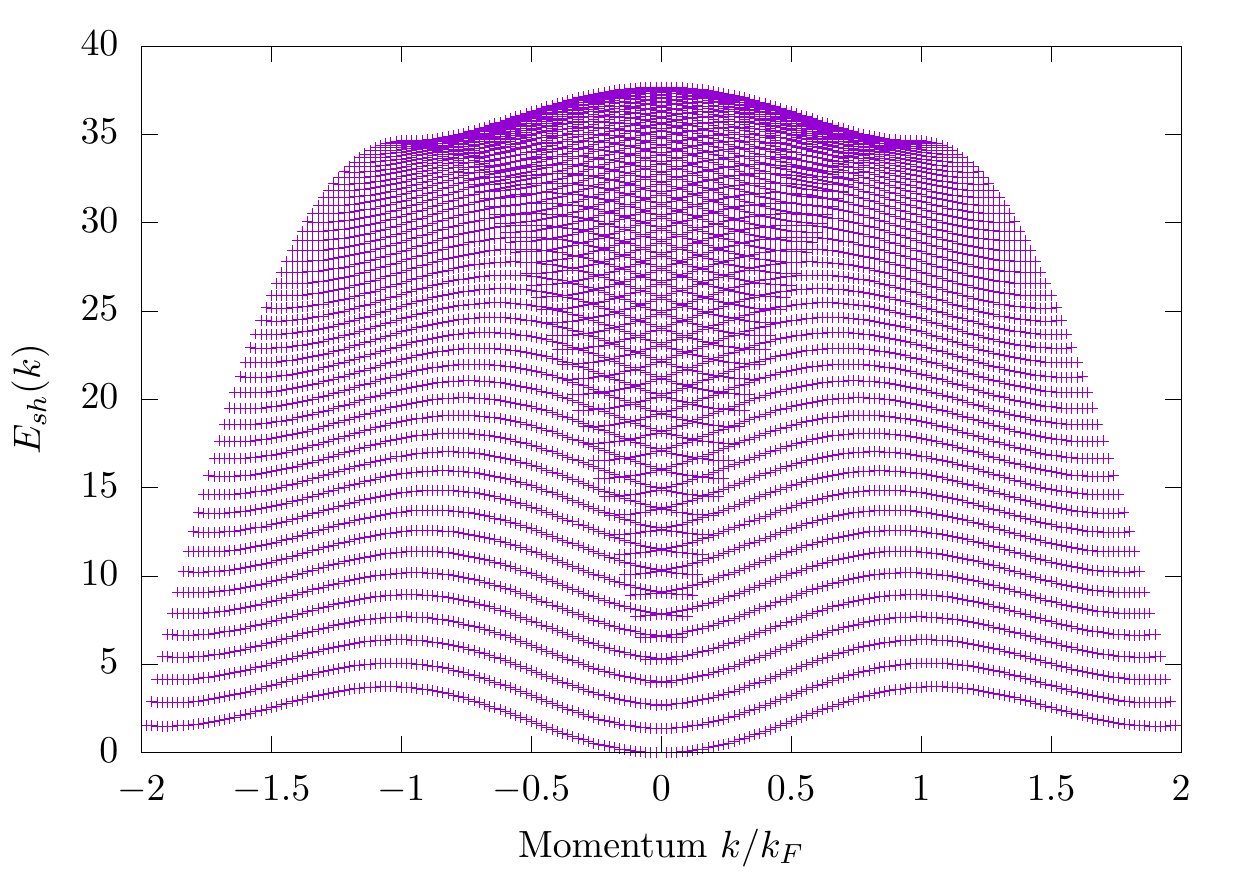}
\end{center}
\caption{The exact spinon-holon ($sh$) two-particle excitation continuum for system size $L=50$ with $\varrho \equiv N/L = 2$ and interaction strength $c=50$.}
\label{Fig:SH}
\end{figure}

We now turn our attention to the spinon-holon two-particle continuum. We compute the energy and momentum of the states with the configurations of integers described in Eqs.~\fr{Eq:IntegersSH} along similar lines to the previous computation. The holon corresponds to a hole in the symmetric Fermi sea with momentum $k_h$; the position of the hole in the Fermi sea of integers is characterized by $0 \leq j \leq N$. The total momentum of the $N$ particle state characterized by integers $j$ and $J$ is
\be
P(j_1,J) = \frac{2\pi}{L}\Bigg( \frac{N}{2} - j - J\Bigg).
\ee
An example of the $sh$ excitation continuum for $N=100$ particles at density $\varrho=2$ is shown in Fig.~\ref{Fig:SH}. There, we have computed the continuum from the numerically exact solution of Eq.~\fr{FullBA1} and Eq.~\fr{FullBA2}. As the spinon excitation is, in fact, a special case of the state under consideration (that with $j = N$), the lower bound of the continuum is given exactly by the spinon dispersion, as is expected, and hence the finite-size effects for the two are identical: the $2k_F$ gap in the $s h$ continuum vanishes in the thermodynamic limit with $M=1$.

It is worth reminding the reader that the $2k_F$ gap is not a result of the finite volume excitation gap: the gap is much larger than the finite volume excitation gap $\delta E_{YG}$~\fr{Eq:LevelSpacingYG} (that is, many small momentum $k\sim0$ spin wave excitations have energy below $\Delta_{2k_F}$). Instead, the non-zero excitation energy at $2k_F$ is a result of interactions between the spinons and holons/antiholons. 

\section{Bound states of spinons: $\Lambda$-strings}

So far, we have focused on elementary excitations of the Yang-Gaudin Bose gas, the spinons and the holons. In the few cases where we have considered more than one spinon, $M>1$, we have been interested in solutions where both spin rapidities are real. However, this is not generically the case: in sectors with $M>1$ the spin rapidities can take complex values, describing multi-spinon bound states.

The complex solutions to the Bethe ansatz equations are organized in regular patterns in the complex plane, known as `strings'. Such solutions were already known by Bethe in his original solution of the Heisenberg model~\cite{BetheZPhys31} and they have been well-studied in spin-$1/2$ quantum magnets (see, for example, Ref.~\cite{EsslerJPhysA92} and the literature review of Ref.~\cite{HagemansJPhysA07}), the Hubbard model~\cite{HubbardBook}, and the attractive Lieb-Liniger model~\cite{TakahashiBook}. They take the form 
\be
\Lambda_\alpha^{n,a} = \omega^n_\alpha + i \frac{c}{2}\Big(n+1-2a\Big)+i\zeta_\alpha^{n,a}, \qquad a = 1,\ldots,n \label{StringWithDevi}
\ee
where $\alpha$ labels the different possible $n$-spinon $\Lambda$-strings, whilst $a$ labels the rapidities contained within the string. The rapidities are distributed symmetrically about the real line, $\Lambda_\alpha^{n,a} = (\Lambda_\alpha^{n,n+1-a})^*$~\cite{VladimirovTheorMathPhys86}, and have the real center $\omega^n_\alpha$. $\zeta_{\alpha}^{n,a} \in \mathbbm{C}$ is known as the string deviation -- in the thermodynamic limit it is assumed to be exponentially small in the system size. The validity of the string hypothesis in the thermodynamic limit of the Yang-Gaudin Bose for computing thermodynamic observables has been confirmed~\cite{KlauserPRA11}, in keeping with general arguments about the structure of the Bethe ansatz equations~\cite{TsvelikAdvPhys83}. We will consider this case in detail and comment on string deviations at the end of this section.  

\subsection{The Bethe-Takahashi equations}
In the thermodynamic limit, the string hypothesis $\zeta_{\alpha}^{n,a} = 0$ is believed to be valid; spin rapidities are organized in regular patterns in the complex plane described by 
\be
\Lambda_\alpha^{n,a} = \lambda^n_\alpha + i\frac{c}{2}\Big(n+1-2a\Big), \qquad a = 1,\ldots,n \label{String}
\ee
which we characterize through their real centers $\lambda^n_\alpha$ and length $n$. Inserting~\fr{String} into the Bethe ansatz equations~\fr{FullBA1},~\fr{FullBA2}, we arrive at the Bethe-Takahashi equations~\cite{KlauserPRA11}
\bea
k_j = 2\pi \frac{\bar I_j}{L} - \frac{1}{L} \sum_{l=1}^N \theta_2(k_j-k_l) + \frac{1}{L} \sum_{n=1}^\infty \sum_{\alpha=1}^{N_n}\theta_n (k_j - \lambda_\alpha^n), \\
\frac{1}{L}\sum_{j=1}^N \theta_n (k_j - \lambda_{\alpha}^n) = 2\pi \frac{\bar J^n_\alpha}{L} + \frac{1}{L} \sum_{m=1}^\infty \sum_{\beta = 1}^{N_m} \Bigg\{ 
(1-\delta_{n,m})\theta_{|n-m|}(\lambda^n_\alpha - \lambda^m_\beta) + 2\theta_{|n-m|+2}(\lambda^n_\alpha - \lambda^m_\beta) \nn
  \hspace{2.5in} + \ldots + \theta_{n+m}(\lambda^n_\alpha - \lambda^m_\beta) \Bigg\}, \label{TBAStringCenter}
\eea
where $\bar I_j \in (\mathbb{Z}+\frac12)$, $\bar J^n_\alpha$ are $N_n$ sets of $n$ numbers in $\mathbb{Z}$ ($\mathbb{Z}+\frac12$) if $N_n$ is even (odd), and we define a convenient reparameterisation of the scattering phase 
\be
\theta_n (u) = - \pi + 2 \arctan\Bigg( \frac{2 u}{c n} \Bigg). 
\ee

\subsection{The Yang-Yang-Takahashi equations}
\label{Sec:YYT}

Now the thermodynamic limit can be taken. Introducing chemical potentials for each of the particle species ($\mu_1$, $\mu_2$) and finite temperature $T$ (we set $k_B = 1$), one can derive the system of Yang-Yang-Takahashi (YYT) equations~\cite{YangJMathPhys69} for the dressed energy $\epsilon(k)$ and the length $n$ $\Lambda$-string dressed energies $\epsilon_n(\lambda)$. A detailed derivation can be found in Ref.~\cite{KlauserPRA11}. The YYT equations read
\bea
\epsilon(k) &=& k^2 - \mu - \Omega  - T \Bigg[ a_2 \ast \ln\Big(1 + e^{-\epsilon/T} \Big) - \sum_{n=1}^\infty a_n \ast \ln\Big(1 + e^{-\epsilon_n/T}\Big)\Bigg](k), \label{YY1}\\
\epsilon_1(\lambda) &=& T\, f\ast \bigg[ \ln\Big( 1 + e^{\epsilon_2/T} \Big) + \ln \Big( 1 + e^{-\epsilon/T} \Big) \bigg] (\lambda), \label{YY2} \\
\epsilon_{n}(\lambda) &=& T\, f\ast \bigg[ \ln\Big( 1 + e^{\epsilon_{n+1}/T} \Big) + \ln \Big( 1 + e^{\epsilon_{n-1}/T} \Big) \bigg](\lambda), \quad n \ge 2,\label{YY3}
\eea
where $\mu = (\mu_1 + \mu_2)/2$, $\Omega = (\mu_1-\mu_2)/2$, and we denote the convolution by
\be
f\ast g(k) = \int_{-\infty}^\infty\rd k'\, f(k-k')g(k') . \label{convol}
\ee
We also define the functions 
\bea
a_n(k) = \frac{1}{\pi} \frac{(nc/2)}{(nc/2)^2 + k^2}, \qquad 
f(k) = \frac{1}{2c} \sech\Big(\frac{\pi}{c}k\Big) . \label{an}
\eea
The YYT equations~\fr{YY1}--\fr{YY3} are supplemented by the asymptotic condition
\be
\lim_{n\to\infty} \frac{\epsilon_n(\lambda)}{n} = 2\Omega. \label{Asymp}
\ee
From this, the large momentum asymptotic form of these functions can be derived~\cite{KlauserPRA11}
\be
\lim_{\lambda\to\infty} \epsilon_n(\lambda) \equiv \epsilon_n^\infty = 2\Omega n + T \ln \Bigg[ \Bigg( \frac{1-e^{-2\Omega(n+1)/T}}{1 - e^{-2\Omega/T}}\Bigg)^2 - e^{-2\Omega n/T}\Bigg]. \label{kinf}
\ee
For $n=1$ (i.e., the spinon) Eq.~\fr{kinf} simplifies to 
\be
\lim_{\lambda \to \infty} \epsilon_1(\lambda) = 2\Omega + T \ln \left[1 + e^{-2\Omega/T} + e^{-4\Omega/T}\right].
\ee
For the case with $SU(2)$ symmetry preserved, $\Omega = 0$, we see that finite temperature $T$ opens a gap
\be
\lim_{\lambda\to\infty}\epsilon_1(\lambda)\Big\vert_{\Omega = 0} = T \ln 3,
\ee
at large spin rapidity (e.g., $k\to2k_F$). This is consistent with the picture presented in Sec.~\ref{Sec:Origin} that finite spinon density (which indeed is induced by finite temperature) leads to a $2k_F$ gap in the spinon dispersion.\footnote{ More generally, with $\Omega = 0$ the length $n$ $\Lambda$-string has a gap of $\lim_{\lambda\to\infty}\epsilon_n(\lambda)\vert_{\Omega = 0} = T \ln [ n(2+n)].$}

The Gibbs free energy per unit length $g$ can be expressed solely in terms of the dressed energy $\epsilon(k)$
\be
g = - \frac{T}{2\pi} \int_{-\infty}^{\infty} \rd k\, \ln \left[ 1+ e^{-\epsilon(k)/T} \right].  \label{Gibbs}
\ee
The density of particles $N/L$ and the spin flip density $M/L$ can be computed by taking derivatives of the Gibbs free energy $g$ with respect to the relevant chemical potentials 
\bea
\frac{N}{L} = - \frac{\p g}{\p \mu}, \qquad \frac{M}{L} = - \frac{1}{2} \Bigg[ \frac{\p g}{\p \mu} - \frac{\p g}{\p \Omega}\Bigg]. 
\eea
A detailed study of the finite-temperature thermodynamics of the Yang-Gaudin Bose gas, including properties such as the polarization of the gas, the specific heat, and local pair correlations, can be found in Ref.~\cite{KlauserPRA11}.

\subsection{Zero temperature limit of the YYT equations}
Let us now consider the zero temperature limit of the YYT equations, on which the remainder of this work will be focused. We begin by defining the momentum $k_F$ for which the dressed energy changes sign $\epsilon(k_F)=0$ (which may be zero in the case of vanishing particle density). Next, we use that the right hand sides of Eqs.~\fr{YY2} and~\fr{YY3} are positive, which implies that $\epsilon_n(\lambda) \geq 0$. It then follows that the YYT equations can be reduced to the form
\bea
\epsilon(k) = k^2 - \mu - \Omega + \int_{-k_F}^{k_F} \rd q\, a_2(k-q)\epsilon(q), \label{YY1T0} \\
\epsilon_1(\lambda) = f\ast \epsilon_2(\lambda) - \int_{-k_F}^{k_F} \rd q\, f(\lambda - q) \epsilon(q), \label{YY2T0}\\
\epsilon_n(\lambda) = f\ast \Big( \epsilon_{n+1} + \epsilon_{n-1} \Big)(\lambda),  \qquad n \geq 2, \label{YY3T0}\\
\lim_{n\to\infty} \frac{\epsilon_n(\lambda)}{n} \to 2 \Omega. \label{limit}
\eea
Notice that Eq.~\fr{YY1T0} has reduced to a Fredholm equation of the second kind, and so it may be solved numerically using standard routines, see~\cite{NR} for one such example.

As is shown in~\ref{App:DressedEnergies}, the solutions to Eqs.~\fr{YY2T0} and~\fr{YY3T0} can be expressed solely in terms of the dressed energy $\epsilon(k)$. The length $n$ string dressed energies satisfy
\bea
\epsilon_n(\lambda) =  2 n \Omega - \Big[a_{n-1}+a_{n+1} \Big] \ast h(\lambda), \label{nstring}
\eea
where we define $a_0(k) = \delta(k)$ and
\bea
h(q) = \int_{-k_F}^{k_F}\rd k\, f(q-k)\epsilon(k). 
\eea

\subsubsection{Strong coupling expansion in the zero temperature limit.} 
To serve as a check of our numerical solution of the YYT equations, we also consider Eqs.~\fr{YY1T0}--\fr{YY3T0} in the strong coupling limit. As previously stated, the dressed energy $\epsilon(k)$ is given by a Fredholm equation of the second kind; it has the formal solution
\bea
\epsilon(k) = k^2 - \mu - \Omega + \int_{-k_F}^{k_F} \rd q\, R(k,q) (q^2 - \mu - \Omega), \\
R(k,q) = \sum_{n=1}^\infty \Big(\frac{1}{\pi c}\Big)^n K_n(k,q), \\
K_n(k,q) = \Bigg( \prod_{i=1}^{n-1} \int_{-k_F}^{k_F} \rd p_i \Bigg) \bar a_2(k-p_1) \bar a_2(p_1-p_2)\ldots \bar a_2(p_{n-1}-q)\ , \\
\bar a_2(x) = \frac{1}{1 + (x/c)^2}\ .
\eea
As this is expressed as a perturbative series, this provides a natural starting point for computing the corrections to the dressed energy $\epsilon(k)$. At zeroth order (e.g., $\pi c=\infty$) have the expected result
\be
\epsilon(k)|_{c=\infty} = k^2 - k_0^2, \label{ecinf}
\ee
where $k_0 = \sqrt{\mu + \Omega}$ is the momentum at which the dispersion relation $\epsilon(k)|_{c=\infty}$ changes sign. Proceeding to order $1/(\pi c)$, we find
\bea
\epsilon(k) &=& \Big( k^2 - k_0^2 \Big) \Bigg\{ 1 + \frac{1}{\pi}\bigg[ \arctan\bigg(\frac{k_F+k}{c}\bigg) + \arctan\bigg(\frac{k_F - k}{c}\bigg)\bigg] \Bigg\} \nn
&& + \frac{c^2}{\pi} \Bigg[ \frac{2k_F}{c}  - \arctan\bigg(\frac{k_F+k}{c}\bigg) - \arctan\bigg(\frac{k_F - k}{c}\bigg)\Bigg] \nn
&& + \frac{ck}{\pi}  \log\Bigg( \frac{c^2+(k_F-k)^2}{c^2+ (k_F+k)^2}\Bigg) + O\Big( (\pi c)^{-2}\Big).
\eea
At small momentum and with a finite particle density (e.g., $k/c \ll 1$ and $k_F/c \ll 1$) this simplifies to  
\bea
\epsilon(k) \approx k^2  - k_0^2 \Bigg( 1 + \frac{4k_0}{3\pi c}\Bigg) + O\Big((\pi c)^{-2}\Big), \label{ksmall}
\eea
where we used $k_F = k_0 + O(c^{-1})$. Equation~\fr{ksmall} can also be obtained by a straight forward perturbative expansion of Eq.~\fr{YY1T0} by insertion of $a_2(x) = 1/\pi c + O(c^{-3})$, valid provided $k_F/c \ll 1$. Notice that this also implies that the effective mass $m_{\rm eff}$ (cf. $\epsilon(k) = E_0 + k^2/2m_{\rm eff}$) is unchanged to $O(c^{-2})$.

Let us now turn our attention to the length $n$-string dressed energies $\epsilon_n(\lambda)$ determined from Eqs.~\fr{YY2T0}~and~\fr{YY3T0}. Using
\be
\int_{-\infty}^{\infty} \frac{ \rd q}{2c}\, \sech\Big(\frac{\pi q}{c}\Big) = \frac12,
\ee
we arrive at the zeroth order expansion
\bea
\epsilon_n(\lambda)|_{c=\infty} = 2n\Omega, \label{cinfen}
\eea
That is, the string dressed energies are equal to their asymptotic values at zeroth order. The $c=\infty$ point is rather simple as the function $f(k)$ becomes infinitely wide, and hence there is no $\lambda$ dependence to the string energies $\epsilon_n(\lambda)$. Physically, this is reasonable: at $c = \infty$ spin flip excitations (e.g., spinons) cannot move through the gas as they are trapped by the hardcore interaction, experiencing Pauli exclusion. Indeed, this is seen as a divergence in the spinon effective mass as $c\to \infty$~\cite{FuchsPRL05,ZvonarevPRL07}.

The next order correction to the length $n$-string dressed energy is considered in detail in~\ref{stringsc}. For $\Lambda$-strings of length $n\ge 2$, we find a remarkably simple expression for the dressed energy
\bea
\epsilon_n(\lambda) = 2n\Omega + \frac{k_F^3}{3\pi} \frac{(nc/2)}{(nc/2)^2 + \lambda^2} + O\Big( (\pi k_F/c)^2 \Big), \qquad n \ge 2, \label{nstringsc}
\eea
whilst the spinon dressed energy is
\be
 \epsilon_1(\lambda) = 2 \Omega + \frac{k_F^3}{6 c \cosh(\pi \lambda/c)} + \frac{k_F^3}{2\pi c} \Bigg[ \psi\Bigg(1 + \frac{i\lambda}{2c}\Bigg) - \psi\Bigg(\frac{1}{2} + \frac{i\lambda}{2c}\Bigg) + {\rm H.c.} \Bigg]
 + O\Big( (\pi k_F/c)^2 \Big),
\ee
where $\psi(x)$ is the digamma function. 

\subsection{Numerical solution of the YYT equations} 

As we want to compute the dispersion relation (e.g., the dressed energy) of $\Lambda$-strings, we need to numerically solve the system of coupled integral equations~(\ref{YY1}--\ref{YY3}). To do so, we follow one of the procedures outlined in Ref.~\cite{KlauserPRA11}:
\begin{enumerate}
\item We begin by introducing a function cutoff $n_{\rm max}$, replacing $\epsilon_{n > n_{\rm max}}$ with their asymptotic value~\fr{Asymp}.
\item We introduce a momentum cutoff $\Delta_i$, reducing the range of the convolution integral~\fr{convol} to $[-\Delta_n, \Delta_n]$. This is quite reasonable as $f(k) \sim e^{-\pi k/c}$ approaches zero exponentially for large $k$, whilst it seems reasonable that $\epsilon(k) \sim k^2$ and the string dressed energies $\epsilon_n(\lambda)$ approaches a constant, cf.~\fr{kinf}. 
\item The solution is now estimated by the set of functions $\{\epsilon, \epsilon_1,\ldots,\epsilon_{n_{\rm max}}\}$ where integrals are evaluated on the truncated range $[-\Delta_i,\Delta_i]$ with $N_i$ evaluation points within this interval.
\item Starting from the non-interacting $c=0$ solution for $\epsilon(k)$ and the asymptotic values~\fr{kinf} for $\epsilon_n(k)$, the solution is then found iteratively.
\item We check convergence by varying $n_{\rm max}$, $\Delta_i$ and the discretization grid $N_i$. 
\end{enumerate}
Alternative numerical approaches, such as using a M\"obius transformation to map the infinite line to the interval $(-1,1)$ (see, for example, Ref.~\cite{PiroliArxiv16}) are tricky to use here as the Kernel $f(k-q)$ becomes close to singular. 

\subsubsection{Results.}
Having discussed both the strong coupling expansion, and our numerical approach to solving the YYT equations, we turn our attention to the results for the dressed energies $\epsilon(k)$ and $\epsilon_n(\lambda)$. 

\begin{figure}[ht]
\begin{center}
\includegraphics[width=0.48\textwidth]{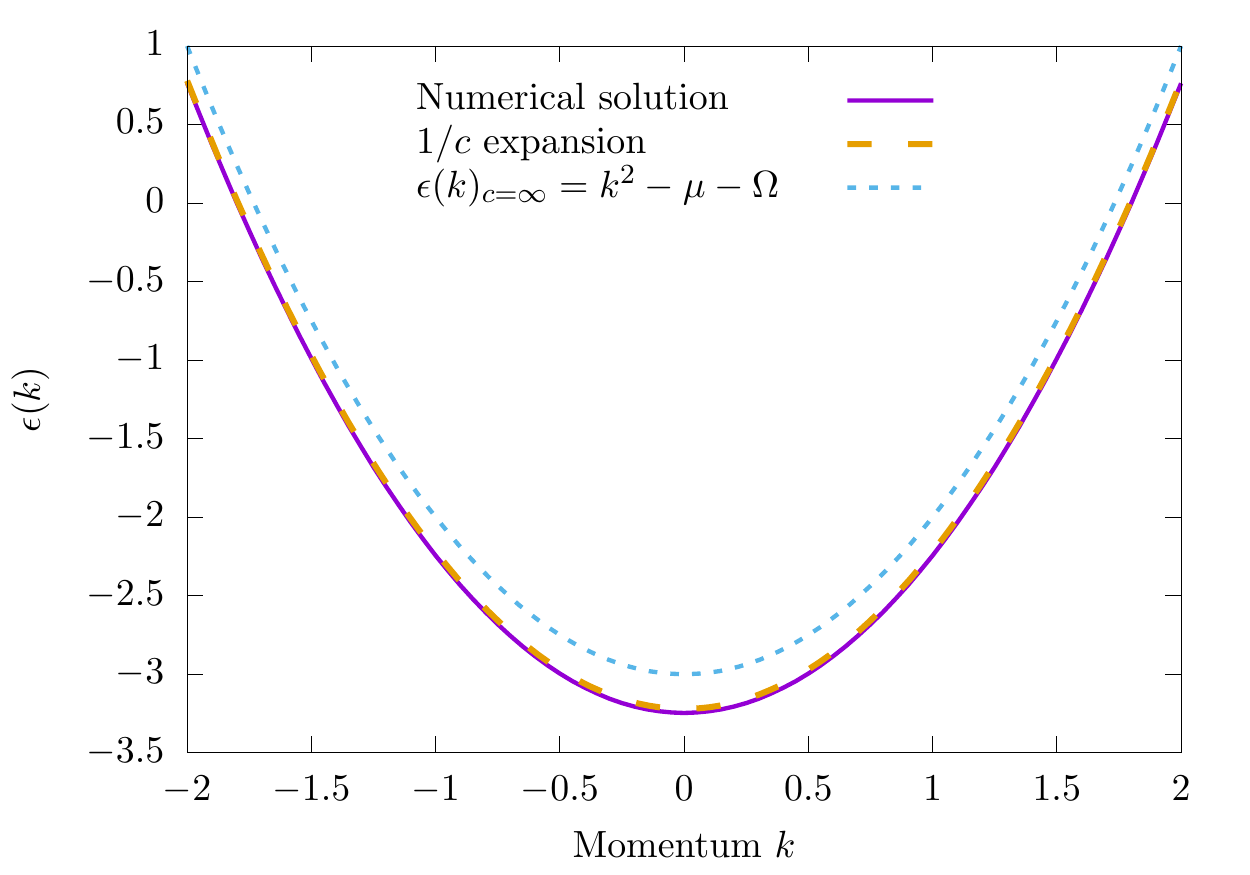}
\end{center}
\caption{The dressed energy $\epsilon(k)$ as a function of momentum $k$ (solid line), obtained by numerical solution of the Yang-Yang-Takahashi equations~\fr{YY1T0}--\fr{YY3T0} with $\mu = 2$, $\Omega = 1$, $c=10$ (solid line). This corresponds to densities of $N/L = 0.6326$ and $M/L \approx 0$. The numerical solution of the integral equations is found with truncation level $n_{\rm max} = 30$, integration range $\Delta_i = 500$ and discretization $\delta k =0.05$. We compare to the strong coupling expansion result~\fr{ksmall} (dashed line) and the hardcore $c=\infty$ result~\fr{ecinf} (dotted line).} 
\label{Fig:DressedE}
\end{figure}

Let us begin with the dressed energy $\epsilon(k)$, from which many thermodynamic properties of the model follow, cf. Eq.~\fr{Gibbs}. In Fig.~\ref{Fig:DressedE} we present a comparison  between the numerical solution of Eq.~\fr{YY1T0} and the strong coupling expansion~\fr{ksmall}, as well as the hardcore $c=\infty$ result~\fr{ecinf}, for the case with $\mu = 2$, $\Omega =1$ and $c=10$.  At small momenta we see that the dressed energy is significantly renormalized to lower energies from the hardcore limit, with the first order expansion~\fr{ksmall} agreeing extremely well with the full numerical solution. For larger momenta (not shown in the plot) the dressed energy $\epsilon(k)$ remains parabolic and close to the $c=\infty$ result. 

We see that the strong coupling result $\epsilon(k) = k^2 - \tilde k_F^2 + O(c^{-2})$ (where $\tilde k_F = k_0[1 + 2k_0/(3\pi c)] \approx k_F$) is a very good approximation for the whole range $k \in [-k_F,k_F]$. As a result, the zero-temperature Gibbs free energy density~\fr{Gibbs} can be evaluated as 
\bea
g &=& \frac{1}{2\pi} \int_{-k_F}^{k_F} \rd k\, (k^2 - \tilde k_F^2) + O(c^{-2}) = -\frac{2}{3\pi}\tilde k_F^3 + O(c^{-2}). 
\eea
Hence we can evaluate, within the $1/c$ expansion, the number density for the different species
\bea
\frac{N}{L} = - \frac{\p g}{\p \mu} = \frac{1}{\pi} \sqrt{\mu + \Omega} + \frac{8}{3\pi^2 c} (\mu + \Omega) + O(c^{-2}), \\
\frac{M}{L} = -\frac12 \Bigg[ \frac{\p g}{\p \mu} - \frac{\p g}{\p \Omega} \Bigg] = O(c^{-2}). 
\eea 
For the parameters in Fig.~\ref{Fig:DressedE} this gives $N/L = 0.6324 + O(c^{-2})$, which is in excellent agreement with the numerically exact result $N/L = 0.6326$. 

\begin{figure}[ht]
\begin{center}
\begin{tabular}{cc}
\includegraphics[width=0.48\textwidth]{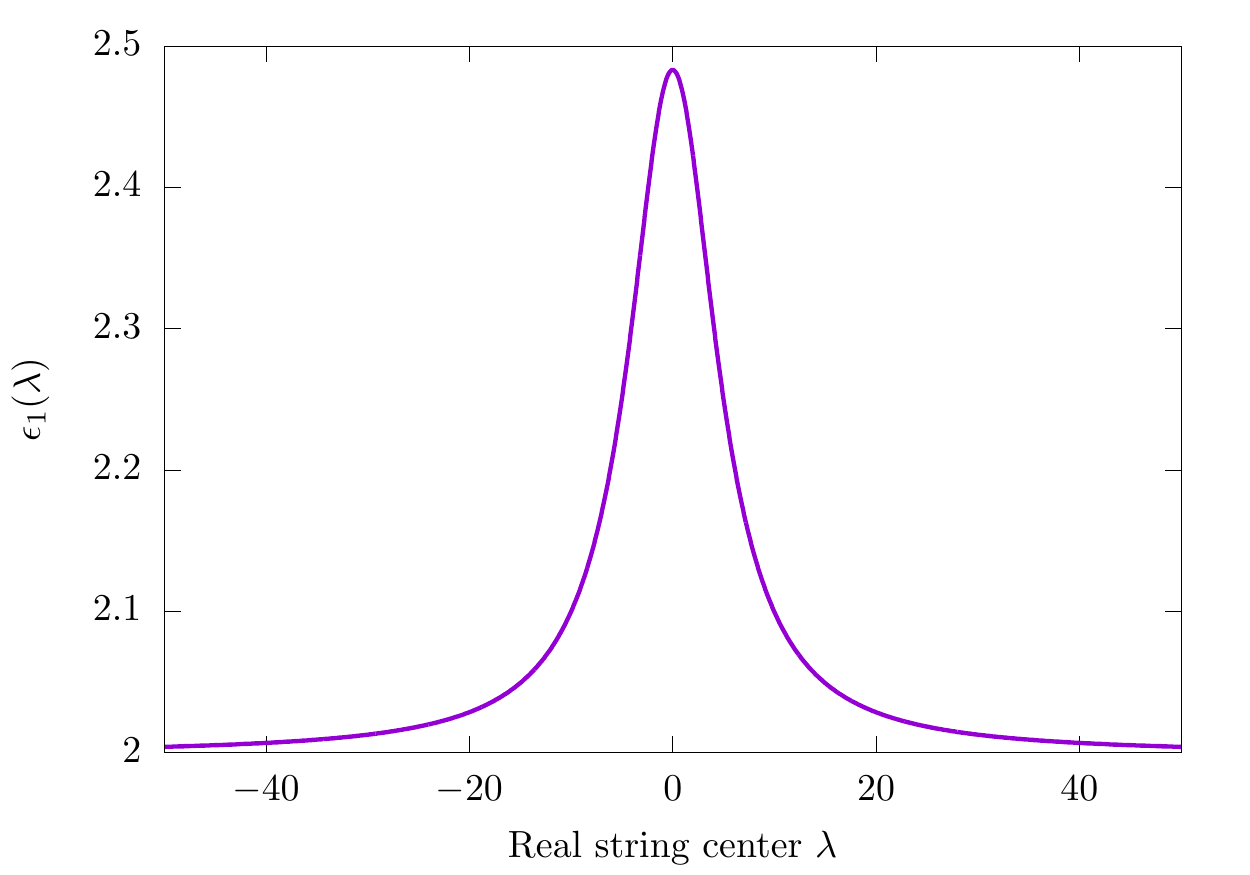} &
\includegraphics[width=0.48\textwidth]{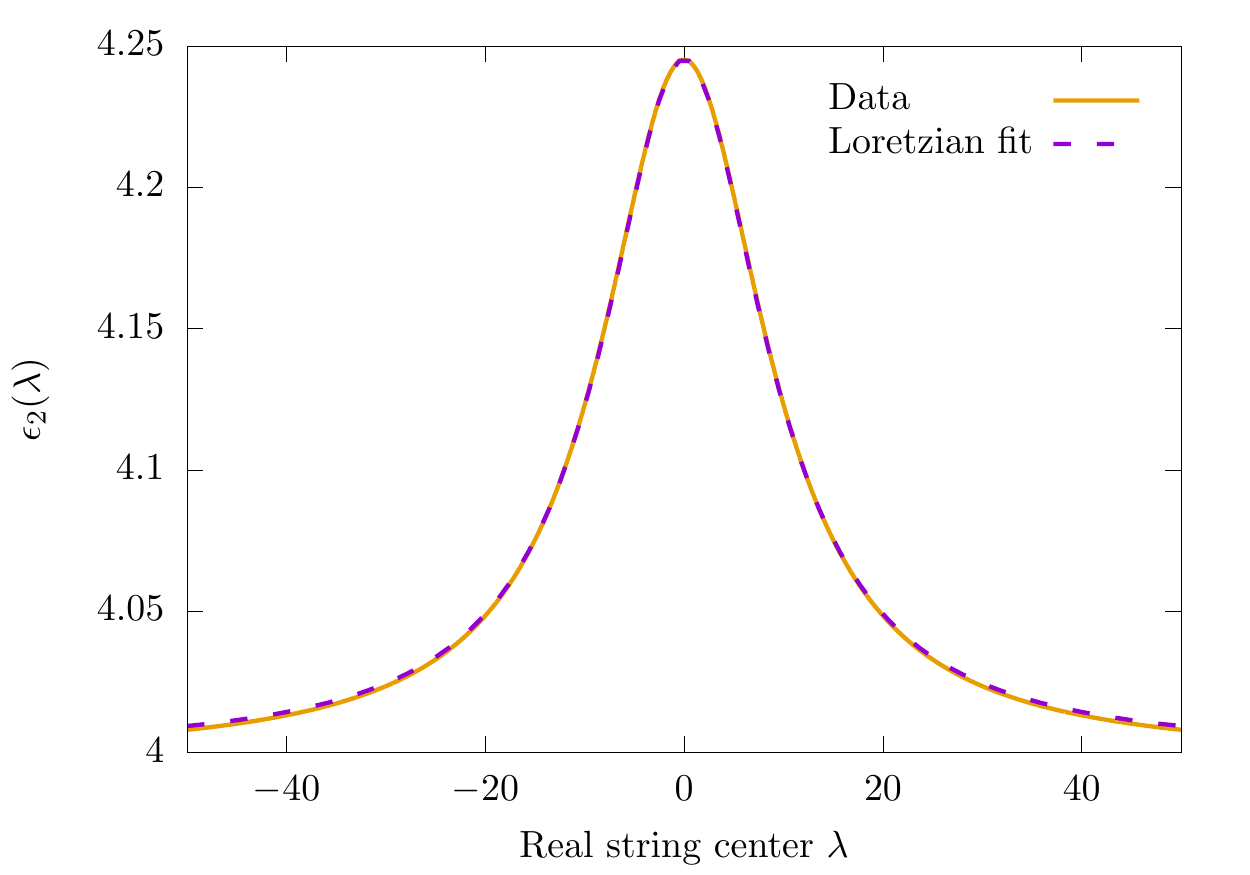}
\end{tabular}
\end{center}
\caption{The length (left panel) $n=1$ and (right panel) $n=2$ string dressed energies $\epsilon_n(\lambda)$ 
plotted against the real string center $\lambda$ for the same parameters as Fig.~\ref{Fig:DressedE}. 
For the $n=2$ string, we show a Lorentzian best fit $f(x) = 4\Omega + a/(c^2 + \lambda^2)$ where $a = 24.536$. The strong coupling expansion~\fr{nstringsc} also predicts a Lorentzian shape, although it significantly underestimates the amplitude (which is not surprising, as the strong coupling expansion parameter $\pi k_F/c > 0.5$ here).} 
\label{Fig:12string}
\end{figure}

\begin{figure}[ht]
\begin{center}
\includegraphics[width=0.48\textwidth]{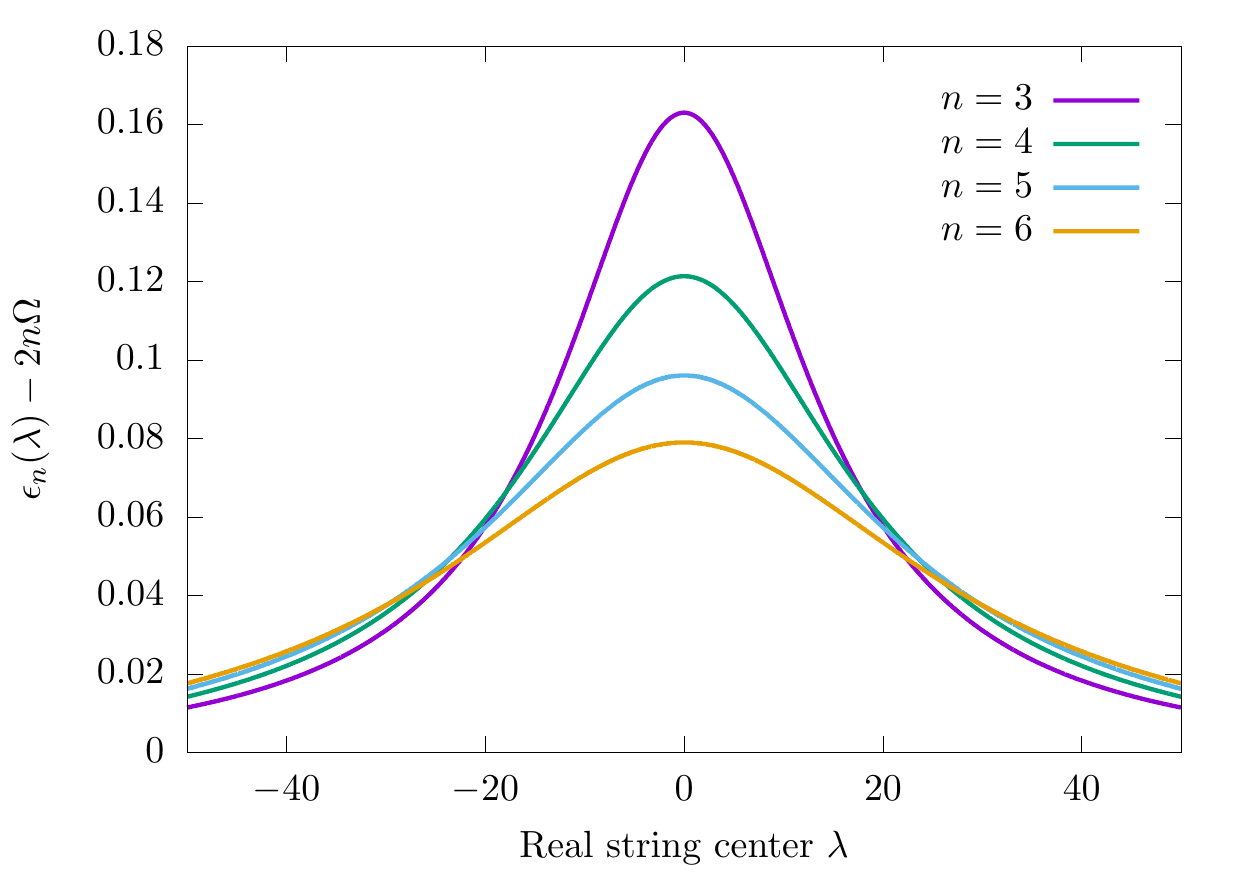}
\end{center}
\caption{As in Fig.~\ref{Fig:12string} for length $n=3-6$ string dressed energies $\epsilon_n(\lambda)$.}
\label{Fig:345string}
\end{figure}

Now, let us turn our attention to the dressed energies $\epsilon_n(\lambda)$, corresponding to the length $n$ $\Lambda$-strings with real center $\lambda$. In Fig.~\ref{Fig:12string} we present the dispersion relations as a function of the real string center $\lambda$ for the length $n=1$ and $n=2$ $\Lambda$-strings, whilst in Fig.~\ref{Fig:345string} we present similar for $n=3-6$. In all cases, we consider the same parameter set as previously, see Fig.~\ref{Fig:DressedE}. The dispersion relations obey the asymptotic condition $\lim_{\lambda\to 0} \epsilon_n(\lambda) = 2n\Omega$,\footnote{We note that the spinon dispersion in Sec.~\ref{Sec:Spinon} is computed for a single spinon in a finite system. Taking the thermodynamic limit $L\to\infty$ with $M=1$ fixed results in $\Omega = 0$. Then we have $\lim_{\lambda \to \infty }\epsilon_1(\lambda) \to 0$, which is consistent with the previously presented results in finite-size systems.} and they are Lorentzian in shape close to $\lambda = 0$ with amplitude decaying as $1/n$. This is consistent with the strong coupling expansion~\fr{nstringsc}, although we note that Eq.~\fr{nstringsc} significantly underestimates the amplitude of Lorentzian for the data presented in Figs.~\ref{Fig:12string} and~\fr{Fig:345string}, which is not too surprising as the strong coupling expansion parameter $\bar k_F = \pi k_F /c > 0.5$ is sizeable.

\subsection{A single 2-string in a finite system}
So far we have studied the thermodynamic limit with finite density of both particle species, which corresponds to a finite density of string excitations of all lengths [cf. Eqs.~(\ref{YY1}-\ref{YY3})].  We now turn our attention to a slightly different scenario, closely related to the cases consider in Secs.~\ref{Sec:Spinon} and~\ref{Sec:TwoParticle}: a state containing a single $2$-string on top of the ground state configuration of integers $\{I_j\}$. We begin by considering small finite systems, in which one needs to take string deviations into account, before considering the large $L$ limit via a strong coupling expansion (see, for example, Secs.~\ref{Sec:Spinon} and~\ref{Sec:TwoParticle}).

\subsubsection{Small systems: string deviations and finite-size effects.}
In a finite size system, the string solutions~\fr{String} are deviated (or deformed) away from their perfect ordering $\Lambda_\alpha^{n,a} \to \Lambda_\alpha^{n,a} + i \zeta_{\alpha}^{n,a}$, where $\zeta_{\alpha}^{n,a} \in \mathbb{C}$. These so-called string deviations have been well studied in integrable lattice models, such as the Heisenberg chain~\cite{TakahashiBook,HagemansJPhysA07,DeguchiJStatMech15}, the Hubbard model~\cite{HubbardBook,DeguchiPhysRep00}, and the spin-1 Babujan-Takhtajan chain~\cite{VlijmJStatMech14}. In continuum integrable models, they have been extensively studied in the attractive limit of the Lieb-Liniger model~\cite{SykesPRA07}. In all of these cases, the string deviations typically vanish exponentially in the system size $L$, and this is expected to be the case in the Yang-Gaudin Bose gas as well~\cite{KlauserPRA11}. 

In this section we will numerically solve the Bethe ansatz equations for small numbers of particles and spinons without assuming the string hypothesis. We limit ourselves to small numbers of particles as this is numerically challenging: we need to solve a system of coupled complex non-linear algebraic equations (there are no good general methods for doing this). In particular, we will focus on the case of the length $n=2$ $\Lambda$-string.

Our aim is to solve the Bethe ansatz equations for $N$ particles with the spin rapidities describing a deviated length $2$-string, parameterized by
\be
\Lambda^{2,\pm} = \lambda \pm \frac{ic}{2} \Big( 1 + 2\delta \Big), \label{2string}
\ee 
where $\lambda, \delta$ are real parameters characterizing the string center and the deviation in the complex plane, respectively. 

We proceed along similar lines to Refs.~\cite{HagemansJPhysA07,VlijmJStatMech14}, by first carefully treating the branch cuts of the arctangent. Working with the branch cut of the logarithm defined by $-\pi < {\rm Im} \ln z \leq \pi$, we have
\be
\arctan(z^*) = \left\{ \begin{array}{lcl} (\arctan z)^* + \pi & \quad & {\rm if~} z\in ]-i,-i\infty[ \\ (\arctan z)^* - \pi & & {\rm if~}z\in]i,i\infty[ \\ (\arctan z)^* & & {\rm otherwise.} \end{array}\right. 
\ee
Taking the sum of the two arctangents with complex conjugate arguments, we have~\cite{HagemansJPhysA07,VlijmJStatMech14}
\be
\arctan(x+iy) + \arctan(x-iy) = \xi(x,1+y) + \xi(x,1-y), 
\ee
with 
\be
\xi(a,b) = \arctan\left(\frac{a}{b}\right) + \pi \Theta(-b){\rm sgn}(a). 
\ee
Here $\Theta(x)$ is the Heaviside function with $\Theta(0) =1$, ${\rm sgn}(a)$ is the signum function with ${\rm sgn}(0)=0$, and we note that the function has the limit $\lim_{b\to0} \xi(a,b) = {\rm sgn}(a)\pi/2 $. 

With these definitions at hand, we can derive the allowed integers $J_1,J_2$ that described the $2$-string.\footnote{When the string hypothesis is assumed, a $2$-string is described by a single integer ($\bar J^n_\alpha$ in Eq.~\fr{TBAStringCenter}) which fixes the real string center $\lambda$. In a finite system, two integers $J_1,\, J_2$ enter the Bethe equations (reflecting that this is a state with two flipped spins) and the string is parameterized in terms of the real string center $\lambda$ and the string deviation $\delta$. We will see that the two integers, $J_1,\, J_2$ are not independent and, indeed, if $\delta = 0$ are fixed to be the same.} Inserting~\fr{2string} into the Bethe ansatz equation~\fr{FullBA2} and then taking the difference between the two equations with integers $J_1,\ J_2$ we have
\bea 
2\pi (J_1-J_2) = \sum_{l=1}^N \Big[ \phi_2(\Lambda^{2,+},k_l) - \phi_2(\Lambda^{2,-},k_l) \Big]+ \phi_1(\Lambda^{2,-},\Lambda^{2,+}) - \phi_1(\Lambda^{2,+},\Lambda^{2,-}).  \label{DiffBAE}
\eea
Matching the real part of the left- and right-hand sides and using ${\rm Re}[\arctan(z) - \arctan(z^*)] = 0$ if ${\rm Re}(z) \neq 0$, we find the following condition on the integers $J_1,J_2$
\bea
(J_2 - J_1) = \left\{ \begin{array}{lll} 1 & \quad & {\rm if~ \delta > 0}, \\ 0 & \quad & {\rm if~} \delta \leq 0. \end{array}\right. \label{diffJ}
\eea
We see that there are two types of string deviations: those which narrow the string in the complex plane $\delta < 0$ (described by $J_1=J_2$), and those which expand it $\delta > 0$ (described by $J_2 = J_1 + 1$). Analogous behavior is seen in other integrable models, see for example Refs.~\cite{HagemansJPhysA07,VlijmJStatMech14}. It is worth noting that once the spin rapidities become complex, we lose the ``Pauli-like exclusion principle'' for the integers $J_1,J_2$.  

Now that we understand the allowed integers $J_1,J_2$ for the spin rapidities, we solve the Bethe ansatz equations obtained by inserting~\fr{2string} into Eqs.~\fr{FullBA1} and \fr{FullBA2} 
\bea
2\pi I_j = k_j L + \sum_{l=1}^N \phi_1(k_j,k_l)- 2 \left[ \xi\Bigg(\frac{2}{c}(k_j - \lambda), -2\delta\Bigg) + \xi\Bigg(\frac{2}{c}(k_j-\lambda),2+2\delta\Bigg)\right], \label{DeviatedBA1} \\
\pi (J_1+J_2) = \sum_{l=1}^N \left[ \xi\Bigg( \frac{2}{c} (\lambda - k_l), 2 + 2\delta\Bigg) + \xi\Bigg( \frac{2}{c}(\lambda-k_l),-2\delta\Bigg) \right],\label{DeviatedBA2} \\
\frac{ \delta^2}{(1+\delta)^2} = \prod_{l=1}^N \frac{\Big(\frac{\lambda-k_l}{c}\Big)^2 + \delta^2}{\Big(\frac{\lambda- k_l}{c}\Big)^2 + (1+\delta)^2}. \label{DeviatedBA3}
\eea
The final equation is equivalent to~\fr{DiffBAE} (it is obtained instead from the non-logarithmic Bethe ansatz equations) and is used for numerical convenience~\cite{VlijmJStatMech14}. We compute the energy and momentum of the $n=2$ $\Lambda$-string relative to the absolute ground state according to 
\bea 
E_{2s} = \sum_{j=1}^N k_j^2 - E_0, \qquad P_{2s} = \sum_{j=1}^N k_j - P_0, \label{2stringEP}
\eea
where $E_0$ ($P_0$) is the ground state energy (momentum). 

\subsubsection{Numerical results.} 
\begin{figure}
\begin{center}
\begin{tabular}{ll}
\includegraphics[width=0.48\textwidth]{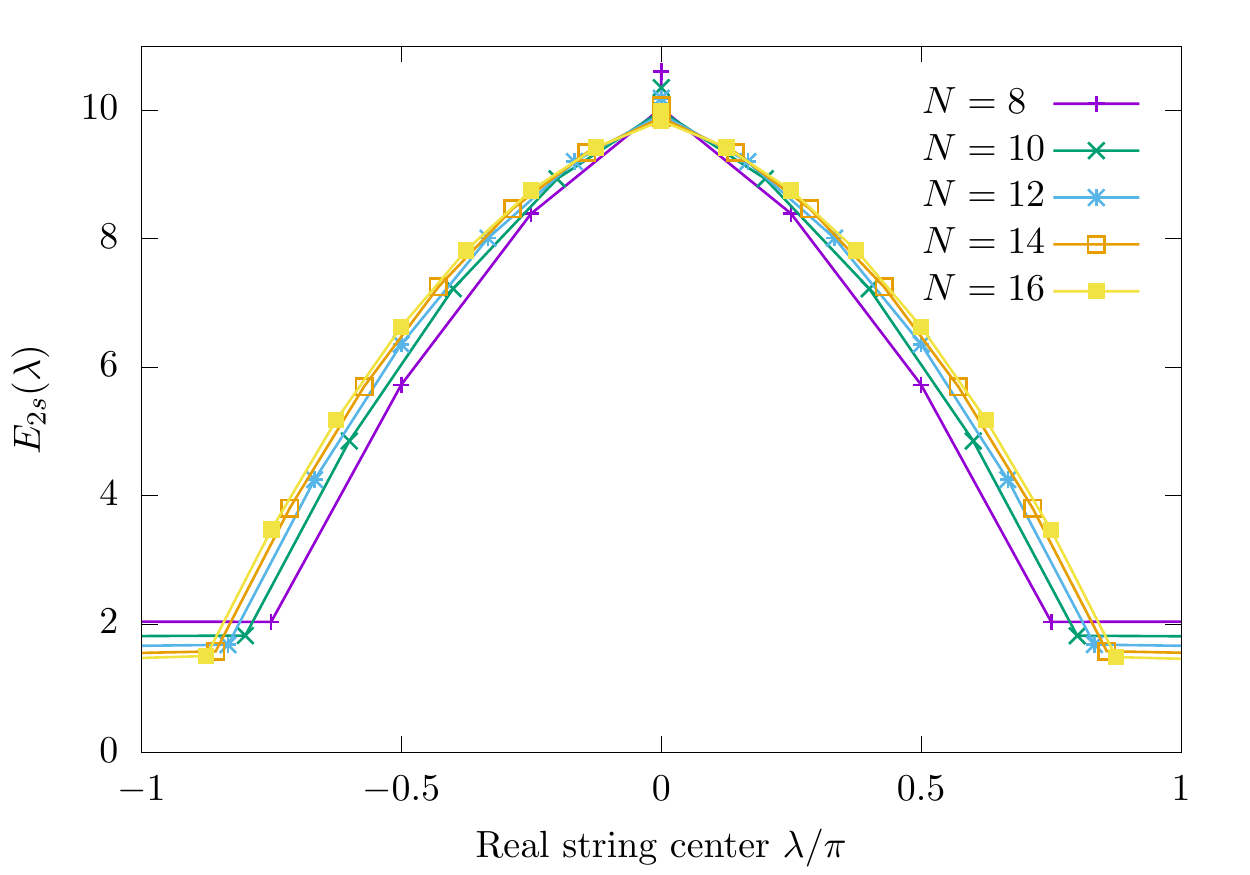} &
\includegraphics[width=0.48\textwidth]{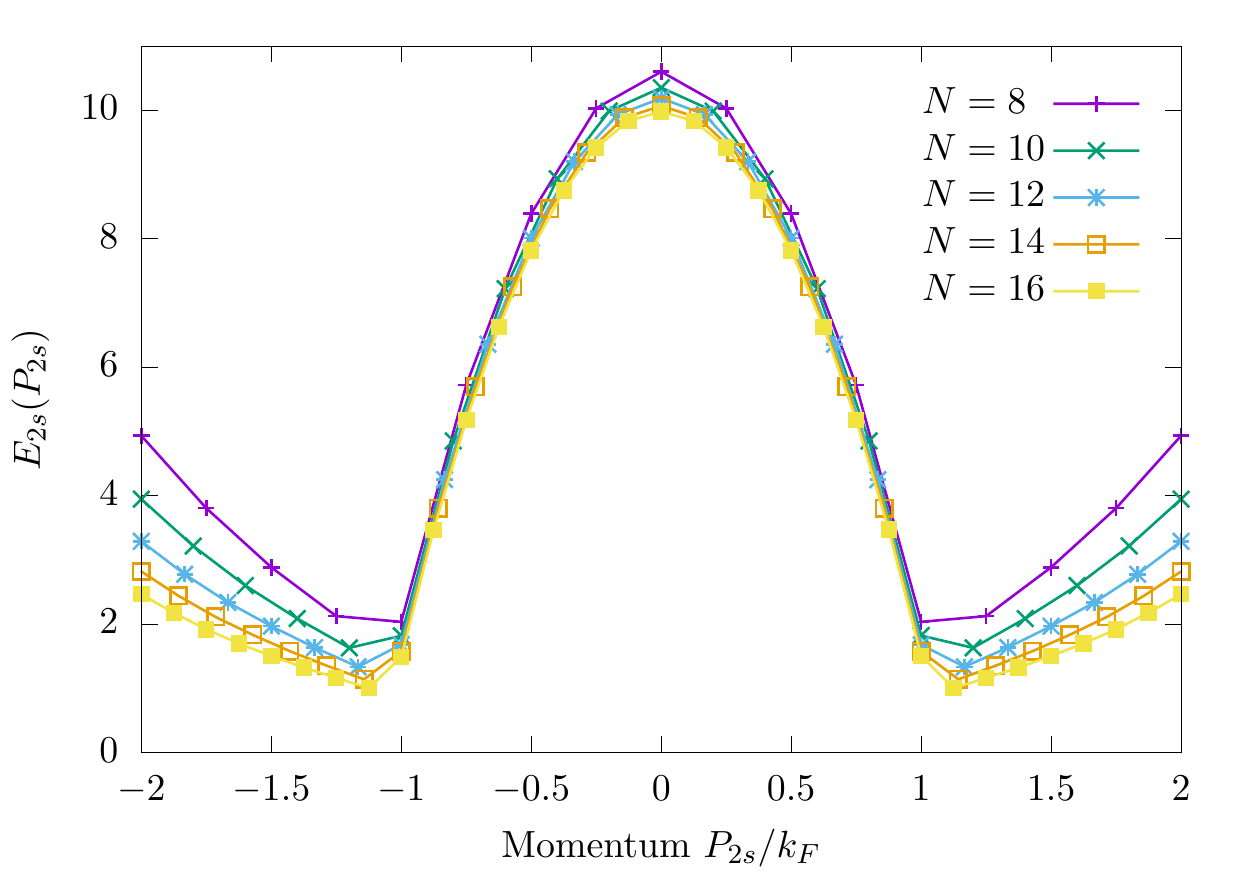} 
\end{tabular}
\end{center}
\caption{The energy of the deviated length $n=2$ $\Lambda$-string as a function of (left panel) the real string center $\lambda$ (close to $\lambda=0$, corresponding to $|P_{2s}| \lesssim k_F$); (right panel) the momentum of the state $P_{2s}$. We consider $N=8-16$ particles at unit filling $\varrho = N/L= 1$ for the interaction strength $c=50$.}
\label{Fig:Deviated}
\end{figure}

We now proceed to solve Eqs.~\fr{DeviatedBA1}--\fr{DeviatedBA3} for the case with the ground state configuration of integers $I_j$~\fr{Eq:IntegersGS}. We vary the integers $J_1,\ J_2$ [subject to the conditions~\fr{diffJ}] to obtain the set of states containing a single $2$-string and we compute the dispersion relation through Eqs.~\fr{2stringEP}. Figure~\ref{Fig:Deviated}(a) shows the dispersion relation as a function of the real string center $\lambda$ for $|\lambda|\leq 1$ (corresponding to $|P_{2s}| \lesssim k_F$), whilst Fig.~\ref{Fig:Deviated}(b) presents the dispersion relation as a function of the $2$-string momentum, $P_{2s}$. 

The structure of the dispersion relation shows two distinct regions of behavior separated by a `kink' at $|P_{2s}|\sim k_F$. This can easily be understood in the following manner. Consider the $2$-string with real string center $\lambda$ in the set of momenta $\{k\}_j$ which is bounded by $k_{R,L} \approx \pm k_F$. In the large system limit, we assume that string deviations are small (i.e., $c \delta \ll |k_j - \lambda|$) and expand the scattering phase between the string center and the momenta as
\be
2 \xi\left( \frac{2}{c}(k_j-\lambda),-2\delta\right) \approx \pi\, {\rm sgn}(k_j-\lambda). \label{approxim}
\ee  
Inserting~\fr{approxim} into~\fr{DeviatedBA1}, we see that the real string center causes an effective shift of the integers $I_j$ according to
\be
I_j \to \left\{ \begin{array}{lll} 
I_j - \frac{1}{2}, & \qquad & {\rm if}~\lambda > k_j, \\
I_j + \frac{1}{2}, & & {\rm if}~\lambda < k_j. 
\end{array}\right. 
\ee
In short, the real string center causes all integers defining momenta to its left to shift to the left, while all those to the right of it shift to the right. This is very much like introducing a hole into the sea of momenta. The kink in the dispersion occurs at $P_{2s} = k_{R,L} \approx \pm k_F$, where the real string center exits/enters the Fermi sea of momenta. Accordingly, in the intermediate region $|P_{2s}| \lesssim k_F$) the dispersion in Fig.~\ref{Fig:Deviated} can be pictured as describing an excitation similar to a holon coupled to a $2$-string.

\begin{figure}
\begin{center}
\includegraphics[width=0.48\textwidth]{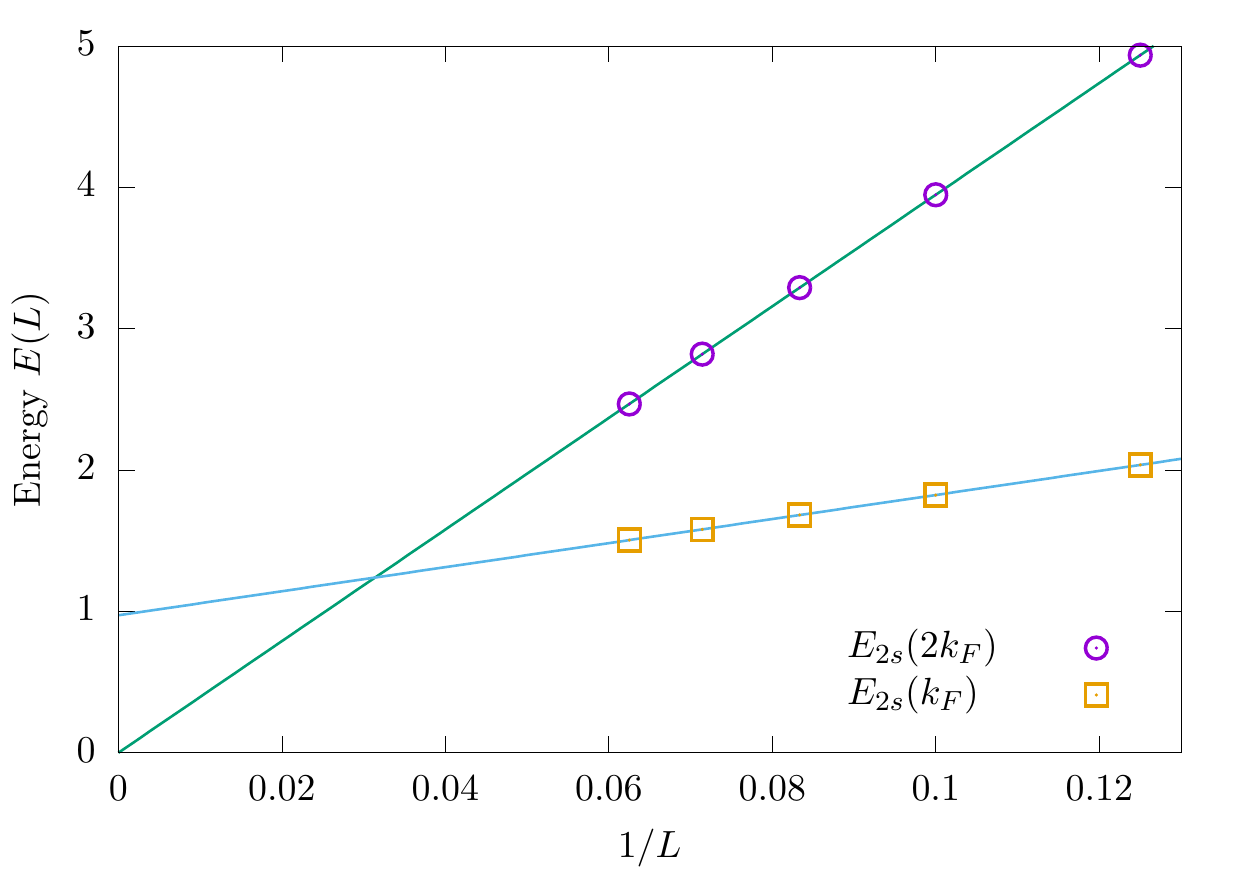}
\end{center}
\caption{Finite size scaling of the energy of the $2$-string with momenta $P_{2s} = k_F, 2k_F$ for the data presented in Fig.~\ref{Fig:Deviated}. Solid lines show $1/L$ fits; the data is consistent with the gap at $2k_F$ vanishing in the $L\to\infty$ limit, whilst the excitation with momentum $k_F$ is of finite energy.}
\label{Fig:2string2kF}
\end{figure}

We see that there are significant finite size effects in the $2$-string dispersion for small numbers of particle $N=8-16$. Qualitatively, we see that there are two different regions: firstly, for $|P_{2s}| < k_F$ the dispersion converges rapidly with increasing particle number. Secondly, the outer region $|P_{2s}| > k_F$ exhibits an upturn that slowly diminishes with increasing system size. Finite size scaling of the $2$-string excitation energies, see Fig.~\ref{Fig:2string2kF}, is consistent with $\lim_{L\to\infty}E_{2s}(2k_F) = 0$, as expected in the thermodynamic limit with $M=2$ (cf. Eq.~\fr{kinf} with $\Omega = 0$). The finite size scaling also shows that the excitation with momentum $P_{2s} = \pm k_F$ remains of finite energy with $L\to\infty$.

\subsubsection{Strong coupling expansion.}
To confirm the structure of dispersion relation for a single 2-string observed in the previous section, we perform a strong coupling expansion along similar lines to Secs.~\ref{Sec:Spinon} and~\ref{Sec:TwoParticle}. In this section we will consider large $L$, adopting the string hypothesis as a working assumption (e.g., we set $\delta \to 0$). The calculation proceeds along similar lines to the spinon case (cf. Sec.~\ref{Sec:Spinon}), resulting in the following equations
\bea
0 = 2 \pi F_{2s}(k_j^{(0)}|\lambda) - \Phi(k^{(0)}_j,\lambda) - \int_{-q_L}^{q_R} \rd q\,  F_{2s}(q|\lambda) K(k_j^{(0)},q) , \\ 
2 \pi (J_1+J_2) = L \int_{-q_L}^{q_R} \rd q\, \rho(q) \Phi(\lambda,q), 
\eea
where $\Phi(u,v) = \phi_1(u,v) + \pi {\rm sgn}(u-v)$, $F_{2s}(k|\lambda)$ is the shift function that characterizes how the presence of the $2$-string with real center $\lambda$ modifies the distribution of momenta $\{k_j\}$ from its ground state configuration $\{k^{(0)}_j\}$, and $J_{1,2} \in \mathbb{Z}+\frac12$ with $J_2 - J_1 = 0,1$ as discussed in the previous section. $q_{R,L}$ denotes the right/left edges of the momenta distribution, which can be determined from the relations~\fr{Eq:NP} with $P = -2\pi (J_1+J_2)/L$, which follows from Eq.~\fr{Eq:P} with $\sum_j I_j = 0$. 

\begin{figure}[h]
\begin{center}
\includegraphics[width=0.48\textwidth]{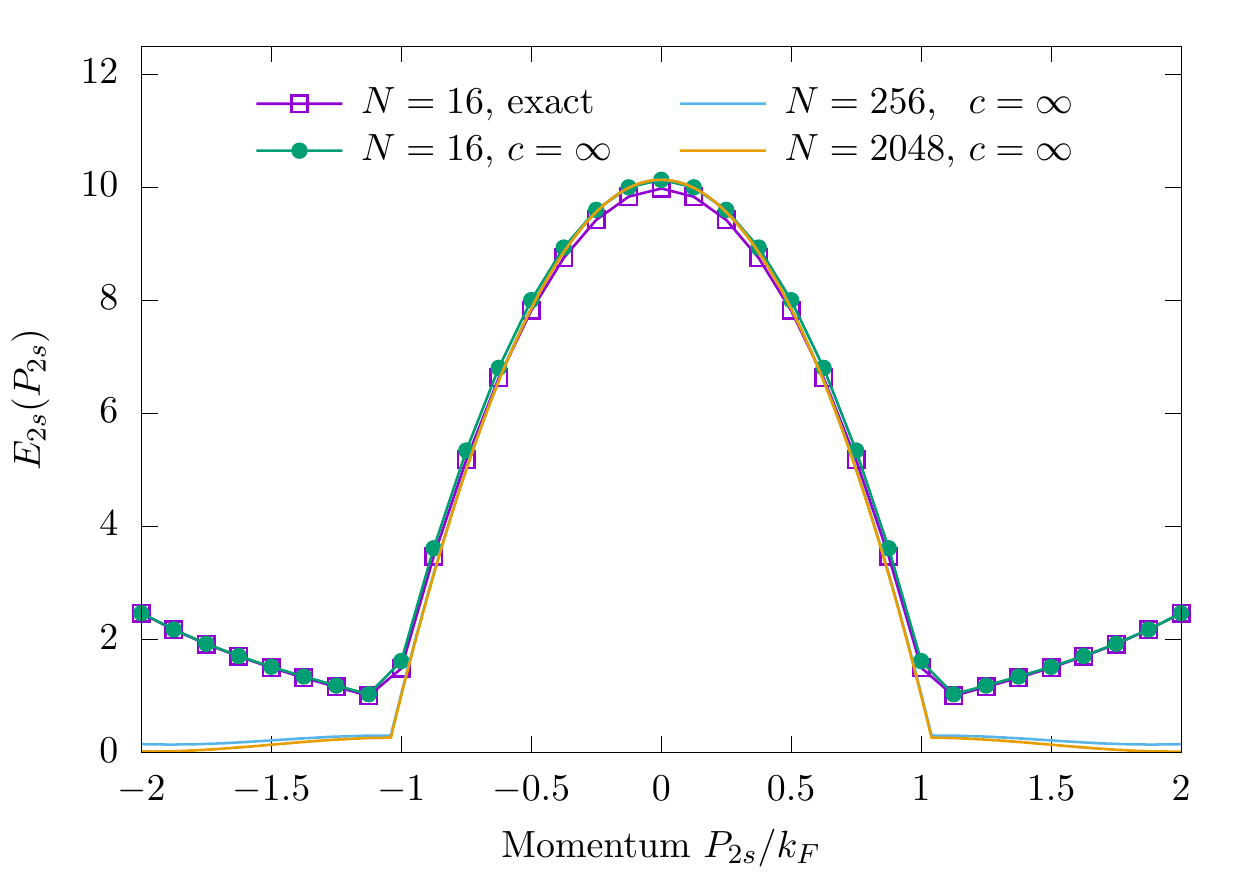}
\end{center}
\caption{A comparison between the strong coupling expansion with $c=\infty$ and the exact result (cf. Fig.~\ref{Fig:Deviated}) for the 2-string dispersion with $N =16$ particles and $c=50$ at unit density. Results for larger systems $N=256,\, 2048$ with $c=\infty$ suggest that, in this limit, the outer region $|P_{2s}|\gtrsim k_F$ becomes increasingly flat as the infinite volume limit is approached.}
\label{Fig:cinf2string}
\end{figure}

The energy and momentum of the $2$-string state with real center $\lambda$ are given by [cf. Eqs.~\fr{Eq:EnuPnu}]
\bea
E_{2s}(\Lambda) = \int_{-q_L}^{q_R} \rd k\, 2 k F_{2s}(k|\Lambda), \qquad P_{2s}(\Lambda) = \int_{-q_L}^{q_R} \rd k\,  F_{2s}(k|\Lambda). 
\eea
We present the $c=\infty$ result in Fig.~\ref{Fig:cinf2string}, comparing to the exact result for the case of $N=16$ particles at unit density with $c=50$. We see excellent agreement between the strong coupling result and the exact result, further confirming the unusual structure of the $2$-string dispersion. This also implies that corrections to the dispersion due to string deviations are rather small, as we take the $\delta\to0$ limit in the strong coupling expansion. We also present $c=\infty$ results for two larger systems, which suggests that in the $N\to\infty$ limit, the $2$-string has a flat (or very close to flat) dispersion when the real string center is located outside the Fermi sea of momenta. This is not entirely surprising: the mass of a single spinon, $m_*$, also diverges in the $c\to\infty$ limit~\cite{FuchsPRL05,ZvonarevPRL07} (see also Sec.~\ref{Sec:Spinon}). 

\section{Discussion}

In this work, we have studied the excitation spectrum of the Yang-Gaudin gas, with a particular focus on how finite system size $L$ affects the dispersion relation for elementary excitations. Using both a strong coupling $1/c$ expansion and exact numerical solution of the Bethe ansatz equations, we have shown that there are prominent finite-size effects in this system, with qualitative and quantitative features changing at energy scales much greater than the finite volume excitation gap. The origin of this is the drastic change in the finite volume excitation gap between the one-component and two-component cases, arising from the new low-energy spin degrees of freedom. 

We started by considering the spinon (spin wave) dispersion above the fully polarized ground state of the two-component gas. It was seen that the spinon dispersion is non-monotonic: at small momentum $k \ll \varrho$, conventional spin wave excitations $E(k) \sim k^2/2m_*$ with effective mass $m_*$ are observed~\cite{FuchsPRL05,ZvonarevPRL07}, whilst at large momentum the energy of the spinon drops, and a roton-like minima occurs close to momentum $|k| = 2k_F$. This minima has finite energy, denoted the roton gap $\Delta_r$, which decreases with increasing system size at fixed density $\Delta_r \sim 1/L$ for  a single spinon $M=1$. For reasonably large systems $L=50-200$, the roton gap is many times greater than the finite volume excitation gap $\Delta_r \gg \delta E_{YG}$. The spinon dispersion also defines the lower bound of the spinon-holon continuum, and accordingly there is a pronounced gap at $k=2k_F$ gap which vanishes as $1/L$ in the system size.  

The origin of the roton-like minima in the spin wave dispersion is the finite interactions between the excitations in the Yang-Gaudin Bose gas. The $2k_F$ gap varies as the product of the spinon density and the particle density. As a result, taking the thermodynamic limit with fixed spinon density ($N$, $M$, $L \to \infty$ with $N/L$, $M/L$ fixed) results in a spinon dispersion with a roton-like minima and $2k_F$ gap. Numerical evidence for this was presented in Fig.~\ref{Fig:MultiSpinon}. The Yang-Gaudin Bose gas thus constitutes an example of roton-like excitations occurring in an exactly solvable microscopic model, a problem which has received attention for over 75 years~\cite{LandauZETF41}--\cite{CohenPR57}. 

The holon-antiholon continuum coincides with that of the (one-component) Lieb-Liniger model. As in that case, there is an energy gap at momentum $2k_F$ which arises from the finite volume excitation gap of the one-component gas, $\delta E_{LL}$, and hence this gap similarly vanishes as $1/L$. However, in the context of the two-component Yang-Gaudin model at strong coupling, we should bear in mind that the energy scale $\delta E_{LL}$ is much greater than the finite volume excitation gap $\delta E_{YG}$ set by the spinon excitations. In other words, unlike the (one-component) Lieb-Liniger model where $\delta E_{LL}$ is a very small energy scale, the $2k_F$ `gap' in the holon-anitholon continuum is actually a large energy scale in the Yang-Gaudin gas (comparable to the roton gap $\Delta_r$) and there are many spinon excitations with energy less than $\delta E_{LL}$, see Eq.~\fr{Eq:LevelSpacingRatio}.

What are the implications of these finite-size effects in the two-component Bose gas? Consider first the finite-temperature properties of the model in a finite-size system. At low temperatures $T \ll \Delta_r$ the system is unaware of the gapped roton-like excitations and hence the low temperature properties are uninfluenced. Raising the temperature, at $T \sim \Delta_r$ the system starts to sample the roton-like minimum and the additional gapped excitations can lead to properties of the system changing -- new excitations with non-zero average momentum are activated and the entropic properties of the system change. It is tempting to suggest that non-monotonicities in physical observables (such as the local density-density correlator) with increasing temperature observed in Ref.~\cite{CauxPRA09} may be due to such excitations becoming thermally activated, but this requires further investigation.

The roton-like minima and anomalously large $2k_F$ gaps in excitation continua may also have profound influences on the non-equilibrium dynamics of the system. Consider the spinon dispersion -- there are now two dominant velocities for excitations: the traditional ($|k| < \varrho$) spin wave velocity $v_s$ and the velocity associated with excitations about the roton-like minima $v_r$. These are generically different in finite-size systems, and their behavior when varying the interaction strength $c$ and the density $N/L$ is different. Accordingly, if a non-equilibrium initial state projects on to the low-lying spinon excitations, one may directly observe (via imaging the local density) the propagation of both types of excitations~\cite{RobinsonArxiv16}. This may provide a root to observing these excitations in experiments on cold atomic gases (see, e.g., Refs.~\cite{PalzerPRL09,CataniPRA12}). The role such roton-like excitations play in the dynamics of a distinguishable impurity immersed in the Bose gas (see, for example, Ref.~\cite{RobinsonPRL16}, where unusual dynamics are observed) is yet to be understood. 

In the second part of work, our attention was turned to the bound state excitations, so-called $\Lambda$-strings, in the Yang-Gaudin Bose gas. These are analogous to well-known excitations in the Heisenberg XXZ model~\cite{TakahashiBook} and the Hubbard model~\cite{HubbardBook}. Focusing on the thermodynamic limit, we solved the Yang-Yang-Takahashi equations numerically and compared results to the strong coupling expansion. We found that the dispersion of $\Lambda$-strings of length $n$, $\epsilon_n(\lambda)$ can be well approximated by a Lorentzian for a wide range of real string centers $\lambda$ about $\lambda = 0$. In the zero temperature limit, this can be understood from the simple relationship~\fr{nstring} between the string energies and the dressed energy $\epsilon(k)$. The relation~\fr{nstring} can be solved under a strong coupling expansion~\fr{nstringsc}, yielding a Lorentzian dressed energy for length $n\ge 2$ $\Lambda$-strings.

We finished by commenting on finite-size effects for the length $n=2$ $\Lambda$-string. To do so, we numerically solved the Bethe ansatz equations for a single $2$-string by treating the string deviations (which are neglected in the thermodynamic limit under the string hypothesis). As with the other single- and multi-particle excitations, we found that there are significant finite size effects in the dispersion of the $2$-strings when system sizes are small; parts of the dispersion relation (those corresponding to the real string center being located outside of the Fermi sea of momenta) converge slowly with increasing system size. Comparison with the strong coupling expansion suggests that corrections to the 2-string dispersion due to the string deviations are very small.

Bound states in integrable models are generally difficult to observe with equilibrium probes, such as inelastic neutron scattering in spin chain materials. However, recently it was realized that bound states can be visible in the non-equilibrium dynamics arising from a local quantum quench~\cite{GanahlPRL12}. The non-equilibrium setting can also allow access to phases which would be thermodynamically unstable in equilibrium, where bound states dominate the physics~\cite{PiroliArxiv16,PiroliPRL16}. Whilst there have been some studies of local quenches in the two-component Bose gas~\cite{RobinsonPRL16,RobinsonArxiv16}, they were in a limit where bound states are absent. Attacking non-equilibrium problems with bound states in multi-component integrable quantum gases remains an outstanding problem.

\ack 
We are grateful to Jean-S\'ebastien Caux, Tam\'as P\'almai, and Eoin Quinn for useful conversations, and Alexei Tsvelik for enthusiastic discussions and encouragement. This work was supported by the Condensed Matter Physics and Materials Science Division, Brookhaven National Laboratory, in turn funded by the U.S. Department of Energy, Office of Basic Energy Sciences, under Contract DE-SC0012704. We acknowledge the hospitality and support of the program \textit{``Mathematical Aspects of Quantum Integrable Models in and out of Equilibrium''} at the Isaac Newton Institute for Mathematical Sciences, University of Cambridge  (under EPSRC grant EP/K032208/1), where this work was initiated. 

\appendix

\section{The continuum limit}
\label{App:limits}
Let us briefly consider taking the continuum limit ($L\to\infty$, $N\to\infty$ with $N/L$ fixed) of Eq.~\fr{Eq:difference}. There, we have the momenta $\tilde k_j$ for a finite momentum state (see the discussion preceding Eq.~\fr{Eq:difference}) and the momenta $k_j$ of the ground state (forming a symmetric Fermi sea with zero total momentum). To take the continuum limit, we use the following identities
\be
\frac1L \sum_l f(\tilde k_l) = \int_{-q_L}^{q_R} \rd \tilde k\, \rho(\tilde k) f(\tilde k),\qquad \frac1L {\sum_l}' f(k_l) = \int_{-q_0}^{q_0} \rd k\, \rho(k) f(k),  \label{Eq:continuum}\\
\ee
where $q_L$ ($q_R$) is the momentum of the left (right) Fermi point for the configuration of integers $\tilde I_0$ and $\tilde J > -N/2$ (e.g., the finite momentum state), whilst $q_0$ is the Fermi momentum of the ground state. We have also introduced the root distributions
\bea
\lim_{L\to\infty} \frac{1}{L(\tilde k_i - \tilde k_{i-1})} = \rho(\tilde k_i), \qquad \lim_{L\to\infty} \frac{1}{L(k_i - k_{i-1})} = \rho(k_i),
\eea
which are identical to leading order in $L$. Notice that this means in the $L\to\infty$ limit, the finite momentum of the state with $\tilde k_j$ is realized through the non-symmetric bounds of the integral $q_L$, $q_R$, which implicitly depend upon $\tilde I_0$ and $\tilde J$.

\section{$\Lambda$-string dressed energies at zero temperature}
\label{App:DressedEnergies}

Let us now consider the string dressed energies which satisfy Eqs.~\fr{YY2T0}~and~\fr{YY3T0}. The integral featuring the dressed energy on the right hand side of~\fr{YY2T0} can be considered as a known function (as the dressed energy can be determined using standard methods for Fredholm equations of the second kind~\cite{NR}). Let us denote this function by
\be
h(\lambda) = \int_{-k_F}^{k_F} \rd q\, f(\lambda-q)\epsilon(q). 
\ee
Motivated by the limit~\fr{limit} and the zeroth order contribution to the strong coupling expansion~\fr{cinfen}, we write the string dressed energy in the form 
\be
\epsilon_n(\lambda) = 2n\Omega + \bar \epsilon_n(\lambda),
\ee 
where $\lim_{n\to\infty} \bar\epsilon_n(\lambda)/n = 0$. The YYT equations~\fr{YY2T0},~\fr{YY3T0} now read
\bea
\bar\epsilon_1(\lambda) &=& f\ast\bar\epsilon_2(\lambda) - h(\lambda), \qquad
\bar\epsilon_n(\lambda) = f\ast\Big(\bar\epsilon_{n+1} + \bar\epsilon_{n-1}\Big)(\lambda), \qquad n \ge 2. 
\eea

We proceed by applying a Fourier transform ${\cal F}$ to the above system of equations. The convolution theorem allows us to decompose the Fourier transform of the convolution into the product of the Fourier transforms
\bea
{\cal F}(f \ast g) = {\cal F}(f) {\cal F}(g), 
\eea 
where we define the Fourier transform as
\bea
{\cal F}(f) \equiv \tilde f(\omega) =  \int_{-\infty}^{\infty} \rd x\, f(x) \exp(-2\pi i \omega x). 
\eea
The YYT equations are now a set of functional equations for the Fourier transform of the dressed energies ${\cal F}(\bar \epsilon_n )= \tilde \epsilon_n(\omega)$: 
\bea
\tilde \epsilon_1(\omega) = - \tilde h(\omega) +  \tilde f(\omega) \tilde \epsilon_2(\omega), \label{YY2FT}\\
\tilde \epsilon_n(\omega) = \tilde f(\omega) \Big[ \tilde \epsilon_{n+1}(\omega) + \tilde \epsilon_{n-1}(\omega) \Big], \qquad n \ge 2 \label{YY3FT}.  
\eea 
Here $\tilde f(\omega) = 1/[2 \cosh(\pi c \omega)]$. We next define the vector $\vec{\varepsilon}(\omega) = (\tilde \epsilon_1(\omega), \tilde\epsilon_2(\omega),\ldots)^{\mathrm T}$ this set of equations can be written as
\bea
\left( \begin{array}{cccccc} -1 & \tilde f(\omega)  &  &  &   \\ \tilde f(\omega)  & -1 & \tilde f(\omega)  &  &   \\  & \tilde f(\omega)  & -1 & \tilde f(\omega)  & &  \\  &  &  \ddots & \ddots & \ddots  \end{array}\right) \vec{\varepsilon}(\omega) 
=   \left( \begin{array}{ccc} \tilde h(\omega) & 0 & \ldots \\ 0 & 0 &  \\ \vdots & &  \end{array} \right). \label{simult}
\eea
Next we truncate the infinite hierarchy of equations at some level $n = n_{\rm max} \gg 1$, and we replace $\epsilon_{n_{\rm max}+1}$ with its asymptotic value. This leads to an additional contribution to the matrix on the RHS of Eq.~\fr{simult} in the lower right corner. Rewriting Eq.~\fr{simult}, we have
\bea
\left( \begin{array}{cccccc} \tilde d(\omega) & 1  &  &  &   \\ 1  &  \tilde d(\omega) & 1  &  &   \\  & 1  & \tilde d(\omega) & 1  & &  \\  &  &  \ddots & \ddots & \ddots  \end{array}\right) \vec{\varepsilon}(\omega) 
=   \left( \begin{array}{cccc} \tilde g(\omega) & 0 & \ldots & \\ 0 & 0 &  & \\ 
\vdots & & \ddots  &  \\
& & & \tilde \epsilon_{n_{\rm max}+1}(\omega)   \end{array} \right), \nn
\label{toinvert}
\eea
where $\tilde g(\omega) = \tilde h(\omega)/\tilde f(\omega)$ and $\tilde d(\omega) = -1/\tilde f(\omega)$.

The solution of the set of equations~\fr{YY2FT}~and~\fr{YY3FT} is now obtained by inverting the matrix on the LHS of Eq.~\fr{toinvert}
\bea
\mathbb{1} \vec{\varepsilon}(\omega) =  R   \left( \begin{array}{cccc} \tilde g(\omega) & 0 & \ldots & \\ 0 & 0 &  & \\ \vdots & & \ddots  &  \\ & & & \tilde \epsilon_{n_{\rm max}+1}(\omega)   \end{array} \right), \quad R = \left( \begin{array}{ccccc} \tilde d(\omega) & 1  &   &   \\ 1  &  \tilde d(\omega) & 1   &   \\  & \ddots  & \ddots & \ddots  &  \\   \end{array}\right)^{-1}. 
\eea
Due to the particularly simple structure of the matrix being inverted, the matrix elements of $R$ are known and have the simple form~\cite{HuJPhysA96}
\be
R_{ij} = - \frac{\cosh[\pi c \omega ( n_{\rm max} + 1 - |j-i| )] - \cosh[\pi c \omega (n_{\rm max} + 1- i - j)]}{2 \sinh(\pi c \omega) \sinh[\pi c \omega (n_{\rm max}+1)]}.
\ee
Ignoring the contribution from the lower right element on the right hand side (which will be justified as we will take $n_{\rm max} \to \infty$), we have
\be
\tilde \epsilon_n(\omega) =  \tilde g(\omega) R_{n1}. 
\ee
We can now take the $n_{\rm max} \to \infty$ limit, which gives
\be
\tilde \epsilon_n(\omega) = - 2 \tilde h(\omega) \cosh(\pi c \omega) e^{-n\pi c |\omega|}. 
\ee
We now find the dressed energy for the length $n$ string by application of the inverse Fourier transform ${\cal F}^{-1}$ and remembering that $\epsilon_n(\lambda) = 2n\Omega + \bar\epsilon_n(\lambda)$
\bea
\epsilon_n(\lambda) = 2 n \Omega +  {\cal F}^{-1}\Big\{ -2 \tilde h(\omega) {\rm sgn}(\omega) \cosh(\pi c \omega) e^{-n \pi c |\omega|} \Big\},
\eea
that is
\be
\epsilon_n(\lambda) =  2 n \Omega -2 \int_{-\infty}^{\infty} \rd \omega\, \tilde h(\omega)\cosh(\pi c \omega)e^{-n\pi c |\omega|} e^{2\pi i \omega \lambda}.
\ee
Substituting in the Fourier transform $\tilde h(q)$ gives
\be
\epsilon_n(\lambda) =  2 n \Omega -2 \int_{-\infty}^{\infty} \rd \omega\int_{-\infty}^\infty \rd q\, h(q)\cosh(\pi c \omega)e^{-n\pi c |\omega|} e^{2\pi i \omega (\lambda-q)}.
\ee

For $n > 1$, computing the $\omega$ integral gives 
\bea
2\int_{0}^{\infty} \rd \omega\, \cosh(\pi c \omega) e^{-n\pi c\omega} \cos\Big[2\pi \omega(\lambda-q)\Big]\nn
\hspace{1cm} = \frac{c(n-1)}{\pi [c^2 (n-1)^2 + 4(\lambda - q)^2]} + \frac{c(n+1)}{\pi [c^2 (n+1)^2 + 4(\lambda-q)^2]} ,
\eea
and hence we have 
\bea
\epsilon_n(\lambda) 
&=& 2 n \Omega - \int_{-\infty}^{\infty} \rd q\, \Big[ a_{n-1}(\lambda-q) + a_{n+1}(\lambda-q) \Big] h(q), \\
&=& 2 n \Omega - [a_{n-1} + a_{n+1}]\ast h(\lambda), \qquad n > 1, \label{Eq:eneq}
\eea
where $a_{n}(k)$ is defined in Eq.~\fr{an}. We see that this expression satisfies the limit conditions $\lim_{n\to\infty}\epsilon_n(\lambda)/n = 2\Omega$ and $\lim_{\lambda\to\infty} \epsilon_n(\lambda) = 2n\Omega$. 

For $n = 1$ we have 
\be
2\int_{0}^{\infty} \rd \omega\, \cosh(\pi c \omega) e^{-\pi c\omega} \cos\Big[2\pi \omega(\lambda-q)\Big] =  \delta(\lambda-q) + \frac{c}{2 \pi [c^2 + (\lambda-q)^2]} ,
\ee
and hence 
\be
\epsilon_1(\lambda) = 2 \Omega - h(\lambda) - a_2 \ast h(\lambda).
\ee

\section{Strong coupling expansion for the string dressed energies}
\label{stringsc}
We begin with Eq.~\fr{nstring}, which was derived in the previous appendix~\fr{Eq:eneq}. The strong coupling solution of this equation is a little more involved, as one cannot take the naive strong coupling expansion of $f(k)$ (due to it featuring in the convolution, so its argument can become arbitrarily large). Instead, we compute $h(q)$ using the strong coupling expression for the dressed energy $\epsilon(k)$ without further approximation. That is, we compute
\bea
h(q) = \int_{-k_F}^{k_F} \rd k\, \frac{k^2-k_F^2}{2c \cosh[\pi(q-k)/c]}, 
\eea
where we use the strong coupling expression for $k_F$. The result is 
\bea
h(q) &=& \frac{q^2 - k_F^2}{\pi} \Bigg[ \arctan\Bigg(\tanh\bigg( \frac{\bar k_F - \bar q}{2}\bigg)\Bigg) + \arctan\Bigg( \tanh\bigg(\frac{\bar k_F + \bar q}{2} \bigg) \Bigg)\Bigg] \nn
&& - \frac{2q}{\pi} \Bigg[ (k_F-q) \arctan\Big( e^{-\bar k_F+ \bar q}\Big) + (k_F+q) \arctan \Big( e^{\bar k_F + \bar q} \Big) \Bigg]\nn
&& + \frac{1}{\pi} \Bigg[ (k_F+q)^2 \arctan\Big( e^{\bar k_F + \bar q} \Big) - (k_F-q)^2 \arctan\Big(e^{-\bar k_F + q}\Big) \Bigg] \nn
&& - \frac{2k_F}{\pi} \left(\frac{c}{\pi}\right) {\rm Im} \Bigg[ {\rm Li}_2\Big(i e^{-\bar k_F + \bar q} \Big) + {\rm Li}_2 \Big( ie^{\bar k_F + \bar q} \Big)\Bigg]\nn
&& - \frac{2}{\pi} \left(\frac{c}{\pi}\right)^2 {\rm Im} \Bigg[ {\rm Li}_3 \Big( i e^{-\bar k_F + \bar q} \Big) - {\rm Li}_3 \Big( e^{\bar k_F + \bar q} \Big) \Bigg]. 
\eea
Here we use the shorthand notation $\bar q = \pi q /c$, $\bar k_F = \pi k_F/c$ and we define the polylogarithm function
\be
{\rm Li}_n(x) = \sum_{k=1}^\infty \frac{x^k}{k^n}. 
\ee
At this point, we perform an expansion in $\bar k_F \ll 1$, which is reasonable for large $c$ and finite particle density. Taking care with the factors of $c/\pi$  which appear in the above expression and expanding the accompanying expression to appropriate orders in $\bar k_F$, we eventually arrive at the simple expression 
\bea
h(q) = - \frac{k_F^3}{6 c \cosh( \pi q/c )} + O( \bar k_F^2). 
\eea 

Now that we have a simple form for $h(q)$, we can compute the convolution
\bea
a_n\ast h(\lambda) &=& - \frac{k^3_F}{6 \pi c} \int_{-\infty}^{\infty} \rd q\, \frac{ (nc/2)}{(nc/2)^2 + (\lambda-q)^2} \frac{1}{\cosh(\pi q/c)}, 
\eea
using the convolution theorem. Fourier transforming the two functions gives 
\be
{\cal F}(a_n) = e^{- n c \pi |\omega|}, \qquad {\cal F}(h) = - \frac{k_F^3}{6} \frac{1}{\cosh( \pi c \omega)}. 
\ee
Multiplying the two Fourier transforms and subsequently taking the inverse Fourier transform yields the convolution
\bea
a_n \ast h(\lambda) = -\frac{k_F^3}{12 \pi c } &\Bigg[& \psi\Bigg( \frac{3}{4} + \frac{n}{4} + \frac{i\lambda}{2c}\Bigg) -  \psi\Bigg( \frac{1}{4} + \frac{n}{4} + \frac{i\lambda}{2c}\Bigg) 
+ {\rm H.c.} \Bigg] 
\eea
where $\psi(x)$ is the digamma function. Now, there is a rather beautiful simplification when we consider the full term appearing in Eq.~\fr{nstring}:
\bea
\Big[ a_{n-1} + a_{n+1} \Big]\ast h(\lambda) &=& - \frac{k_F^3}{12 \pi c} \Bigg[ \psi\Bigg(1 + \frac{n}{4} + \frac{i\lambda}{2c}\Bigg) - \psi\Bigg( \frac{n}{4} + \frac{i\lambda}{2c} \Bigg)+ {\rm H.c.} \Bigg],\nn
&=& - \frac{k_F^3}{3 \pi } \frac{(nc/2)}{(nc/2)^2 + \lambda^2}, 
\eea
which follows from the property of the digamma function $\psi(1+z) = \psi(z) + 1/z$. So, we arrive at the simple expression
\bea
\epsilon_n(\lambda) = 2 n \Omega + \frac{k_F^3}{3 \pi} \frac{(nc/2)}{(nc/2)^2 + \lambda^2}+ O(\bar k_F^2), \qquad n > 1. 
\eea
For the special case of $n=1$, this simplification does not occur and we have, explicitly,
\bea
\epsilon_1(\lambda) &=& 2 \Omega + \frac{k_F^3}{6 c \cosh(\pi \lambda/c)} + \frac{k_F^3}{2\pi c} \Bigg[ \psi\Bigg(1 + \frac{i\lambda}{2c}\Bigg) - \psi\Bigg(\frac{1}{2} + \frac{i\lambda}{2c}\Bigg) + {\rm H.c.} \Bigg]
+ O(\bar k_F^2).
\eea

\end{document}